\documentclass[]{aa}
\usepackage{graphicx}
\usepackage{subcaption}
\usepackage{txfonts}
\usepackage{natbib}
\usepackage{longtable}
\usepackage{xspace}
\bibpunct{(}{)}{;}{a}{}{,}   % A&A style
\usepackage{color}

\newcommand{\magpt}[2]{%
\mbox{$\rm #1\hspace{-0.25em}\stackrel{m}{.}
\hspace{-1.0mm}#2$}}  % magnitude \magpt {}{}

\newcommand{\halpha}{H\ensuremath{\alpha}\xspace}
\newcommand{\hbeta}{H\ensuremath{\beta}\xspace}
\newcommand{\heps}{H\ensuremath{\epsilon}\xspace}

\newcommand{\msun}{\ensuremath{\mathrm{M}_\odot}}
\newcommand{\kms}{\ensuremath{\mathrm{km}\,\mathrm{s}^{-1}}}

\newcommand{\teff}{\ensuremath{T_{\mathrm{eff}}}}

\newcommand{\vrot}{\ensuremath{v_{\mathrm{rot}}}}
\newcommand{\ggll}{G\&L}
\newcommand{\poly}{polyn.}

\newcommand{\fitsb}{{\sc fitsb2}\xspace}

\newcommand{\scq}[1]{{\color{blue}[SC, Q: \emph{#1}]}}
\renewcommand{\scq}[1]{}

\newcounter{trmc}
\newcommand{\trm}[1]{\addtocounter{trmc}{1}{\color{blue}[TRM~\arabic{trmc}: \emph{#1}]}}
\renewcommand{\trm}[1]{}

\newcounter{rnc}
\newcommand{\rn}[1]{\addtocounter{rnc}{1}{\color{green}[RN~\arabic{rnc}: \emph{#1}]}}
\renewcommand{\rn}[1]{}

\newcommand{\mtot}{M_{\mathrm{tot}}}
\newcommand{\mch}{M_{\mathrm{Ch}}}

\def\apgt{{\raise-.5ex\hbox{$\buildrel>\over\sim\,$}}}
\def\aplt{{\raise-.5ex\hbox{$\buildrel<\over\sim\,$}}}
\begin{document}

\title{The ESO supernovae type Ia progenitor survey (SPY)%
\thanks{Based on data obtained at the Paranal Observatory of the European 
Southern Observatory for programs No. 165.H-0588, 167.D-0407, 71.D-0383,
72.D-0487}
}

\subtitle{The radial velocities of 643 DA white dwarfs}

\author{R.~Napiwotzki\inst{1,2}
\and C.A.~Karl\inst{2}
\and T.~Lisker\inst{2,3}
\and S.~Catal\'an\inst{4}
\and H.~Drechsel\inst{2}
\and U.~Heber\inst{2}
\and D.~Homeier\inst{5}
\and D.~Koester\inst{6}
\and B.~Leibundgut\inst{7}
\and T.R.~Marsh\inst{4}
\and S.~Moehler\inst{7}
\and G.~Nelemans\inst{8}
\and D.~Reimers\inst{9} 
\and A.~Renzini\inst{10}
\and A.~Str\"oer\inst{2}
\and L.~Yungelson\inst{11} 
}

\institute{Centre for Astrophysics Research, University of 
  Hertfordshire, Hatfield, AL10 9AB, UK\\
  \email{r.napiwotzki@herts.ac.uk}
  \and
  Dr. Remeis-Sternwarte \& ECAP, Universit\"at Erlangen-N\"urnberg,
  Sternwartstr. 7, D-96049 Bamberg, Germany
  \and 
  Astronomisches Rechen-Institut, Zentrum f\"ur Astronomie der Universit\"at
  Heidelberg, M\"onchhofstr. 12-14, 69120 Heidelberg, Germany
  \and
  Department of Physics, University of Warwick, Coventry CV4 7AL, UK
  \and
  Georg-August-Universit\"at, Institut f\"ur Astrophysik,
  Friedrich-Hund-Platz 1, 37077, G\"ottingen, Germany
  \and
  Institut f\"ur Theoretische Physik und Astrophysik, 
  Universit\"at Kiel, D-24098 Kiel, Germany
  \and
  European Southern Observatory, Karl-Schwarzschild-Str. 2,
  D-85748 Garching, Germany
  \and
  Institute for Astronomy, KU Leuven, Celestijnenlaan 200D, B-3001 Leuven,
  Belgium
  \and
  Hamburger Sternwarte, Gojenbergsweg 112, 21029, Hamburg, Germany
  \and
  INAF - Osservatorio Astronomico di Padova, vicolo dell'Osservatorio 5,
  35122, Padova, Italy
  \and
  Institute of Astronomy of the Russian Academy of Sciences, 48 
  Pyatnitskaya Str., 119017 Moscow, Russia
}

\date{Received ??;  accepted ??}

\abstract {Close double degenerate binaries are one of the favoured progenitor
  channels for type Ia supernovae, but it is unclear how many suitable systems
  there are in the Galaxy.  We report results of a large radial velocity
  survey for double degenerate (DD) binaries using the UVES spectrograph at
  the ESO VLT (ESO \textbf{S}N\,Ia \textbf{P}rogenitor surve\textbf{Y} --
  SPY). Exposures taken at different epochs are checked for radial velocity
  shifts indicating close binary systems.  We observed 689 targets classified
  as DA (displaying hydrogen-rich atmospheres), of which 46 turned out to
  possess a cool companion. We measured radial velocities (RV) of the
  remaining 643 DA white dwarfs.  
  We managed to secure observations at two or
  more epochs for 625 targets, supplemented by eleven objects meeting our  selection criteria from literature. The data reduction and analysis methods applied to the survey data are described in detail. The sample contains 39 double
  degenerate binaries, only four of which were previously known. 20 are
  double-lined systems, in which features from both components are visible,
  the other 19 are single-lined binaries. We provide absolute RVs transformed
  to the heliocentric system suitable for kinematic studies. Our sample is large enough to sub-divide by mass: 16 out of 44 low mass targets ($\le 0.45\,\msun$)
  are detected as DDs, while just 23 of the remaining 567 with multiple spectra and mass $> 0.45\,\msun$ are double. Although the detected fraction amongst the low
  mass objects ($36.4 \pm 7.3$\%) is significantly higher than for the higher-mass, carbon/oxygen-core dominated part of the sample ($3.9 \pm 0.8$\%), it is lower than the detection efficiency based
  upon companion star masses $\ge 0.05\,\msun$. This suggests either companion stars of mass $< 0.05\,\msun$, or that some of the 
  low mass white dwarfs are single.
  }


%\authorrunning{Napiwotzki et al.}
% \titlerunning{The ESO supernovae type Ia progenitor survey (SPY). I.}

\maketitle

\section{Introduction} 

\label{intro}

%\subsection{Background}
%\label{sec:backg}

Type~Ia supernovae (SNe\,Ia) play an outstanding role in the study of cosmic
evolution, for instance, they served as standardizable candles in the studies 
that led to the discovery of the acceleration of the Universe's
expansion \citep[e.g.][]{1998AJ....116.1009R, 1999ApJ...517..565P}.  There is
general consensus that SNe\,Ia events are due to thermonuclear explosions
in degenerate matter \citep{1960ApJ...132..565H}.  However, the nature of
their progenitors, as well as explosion mechanisms and origin of different 
subclasses still pose  serious problems.

\citeauthor{1960ApJ...132..565H} suggested that explosion occurs when a critical
mass (likely the Chandrasekhar limit, $1.4\,\msun$) is reached. The most natural candidates
for explosions are then white dwarfs in binaries accreting matter from
non-degenerate hydrogen companions via stable Roche-lobe overflow
\citep{1963stev.conf..389S,1971Ap&SS..11..373W, 1973ApJ...186.1007W} or stellar
wind capture \citep{1971Ap&SS..14..179T,1976Ap.....12..342T}. This is the earliest 
``classical'' so-called single-degenerate (SD) scenario. 

Another ``classical'' scenario is the ``double-degenerate'' (DD) one, in which
explosion is, hypothetically, the outcome of the merger of two white dwarfs of 
(super)-Chandrasekhar total mass \citep{1979wdvd.coll..426W,
1979AcA....29..665T, 1981NInfo..49....3T,1984ApJ...277..355W,1984ApJS...54..335I}. 
The merger of components is due to the orbital shrinkage because of the loss of angular momentum via gravitational wave radiation and subsequent unstable mass loss. 
Initially, it was envisioned that the lighter of two white dwarfs fills its Roche lobe first 
since  $R_{\mathrm{WD}} \propto M_{\mathrm{WD}}^{-1/3}$ and, by virtue of the same
$M-R$ relation, in systems with mass
ratios $\gtrsim2/3$, the merger occurs on the dynamical time scale
(comparable to few orbital periods).  The disrupted white dwarf transforms
into a ``heavy disc'' or ``envelope'' from which the matter accretes onto the
central object. This inference was confirmed by Smooth Particle Hydrodynamics (SPH) calculations by
\citet{1990ApJ...348..647B} and in many other theoretical studies. 

Later, it was realised that the stability of the mass transfer depends also on the
efficiency of spin-orbit coupling, which affects the variation of
accretion rate as the merger takes place \citep{2004MNRAS.350..113M}.
SPH-calculations using a physical equation of state, showed
that the lighter component disrupts within about 100 orbital periods and the merger may
proceed either via a disk or direct impact, depending on the mass
ratio of the components \citep{souza_dynme06,2009JPhCS.172a2034D}.
 
Currently, two possible DD scenarios are commonly discussed. In ``violent'' (a.k.a. ``prompt'') mergers the matter of the
lighter WD, may, even before its complete disruption, start to accrete onto the
surface of the companion on a dynamical timescale. Detonation is supposed to be
initiated in a He+C+O mixture at the interface of the merging white dwarfs.
% and to
%result in complete burning of both white dwarfs.
Helium detonation produces a shock wave, which propagates and focusses towards the center of the accretor and results in ignition of carbon and explosion 
\citep{1990ApJ...354L..53L,lg91}. 

If the explosion does not occur promptly (at lower mass ratios than required 
for violent merger, see, e.g., discussion by \citet{2016ApJ...821...67S}),	
an object consisting of a cold virtually
isothermal core, a pressure-supported envelope, a Keplerian disc, and a tidal tail
forms \citep{guerrero04}. A
He or C+O-mixture may explode in the envelope which is
the hottest part of the object 
and lead to the detonation of carbon at the periphery of the accretor.
If two C/O white dwarfs gradually merge, depending on the rate of settling of the envelope matter onto the core, either central ignition of carbon and a SN~Ia happens or a neutron star 
or a massive Oxygen-Neon (O/Ne) white dwarf  forms \citep{nomoto_iben85,1990A&A...236..378M}.

In both varieties of the DD scenario, explosion of the C/O core, which may have a sub-$\mch$\ mass, results in the
disruption of the entire configuration. The major fraction of the accretor mass
burns to radioactive Ni, which determines the optical luminosity and the spectrum of
the SN, while the donor burns to intermediate mass elements which define
the observational manifestations of SNe at maximum brightness
\citep{1992ApJ...386L..13S,2010ApJ...714L..52S}.

An advantage of the DD scenario is that it produces a delay-time distribution
(DTD) which overlaps with the Hubble time, while in the SD channel delays are shorter
than $\simeq3$~Gyr \citep[see, e.
g.,][]{2010AstL...36..780Y,2013A+A...552A..24B}. If mergers of carbon/oxygen core white dwarf pairs (C/O+C/O), as
well as mergers of C/O~white dwarf and massive He and hybrid (C/O core + thick He mantle)
white dwarfs are taken into account, population synthesis studies can reasonably 
reproduce the current Galactic SNe~Ia rate as well as the slope of their observationally 
inferred  delay time distribution \citep[e.g., ][]{2017MNRAS.464.1607Y}.

We refer the interested reader for further details concerning evolution
of close binaries leading to the formation of progenitors of SNe~Ia and the physical 
processes involved in explosions of SN~Ia and their observational 
manifestations to a number of recent reviews
\citep[e.g.][and references therein]{2013FrPhy...8..116H,2014LRR....17....3P,
2014ARA+A..52..107M,2014NewAR..62...15R,
2017suex.book.....B,2017hsn..book..317T,2018PhR...736....1L,
2018SSRv..214...72R}.

The significance of the DD scenario relies upon the existence in nature of a
sufficient number of DDs prone to merge and the physical parameters of their
components. A given Galactic SN Ia rate and pre-supernova lifetime implies a
specific number of progenitors in the Galaxy. Searches for DDs thus test
them as progenitors of Type~Ia SNe, but also provide benchmarks for testing
the theory of binary star evolution, since in the process of formation, DDs
pass through two to four very badly understood stages of common envelope
evolution.

The first known DD, L870-2, was discovered by chance in the 1980s
\citep{1988ApJ...334..947S}, but its orbital period of $1.55\,$days is too
long for it to merge within a Hubble time and thus be representative of the
putative DD progenitors of present-day Type~Ia supernovae, which must have
periods of less than $10$ to $15\,$hours. Systematic surveys of
the same decade \citep{1987ApJ...322..296R, 1991ApJ...374..281F}
failed to find any DDs among 44 and 25 targets, respectively.  
A sure-fire DD as well as some candidates were found by \cite{1990ApJ...365L..13B} 
among 54 objects. The (premature) conclusion of these studies was that the space density of DDs 
was below that required for reproducing the observed rate of SNe~Ia. The 
low masses ($\mtot < \mch$) of the confirmed and candidate DDs as well as 
their long merger timescales were raised as further arguments against 
the DD scenario.

In the 1990s, more successful targeted searches were made, focusing upon low
mass (helium) white dwarfs, which primarily form via binary evolution
\citep{1995MNRAS.275L...1M,1995MNRAS.275..828M,1997MNRAS.288..538M,
2000MNRAS.314..334M}. By the year 2000, around $180$ white dwarfs had been checked
for RV variations yielding a sample of 18 DDs with $P<6\fd 3$
\citep{2000NewAR..44..119M, 2000MNRAS.319..305M}. None of the 18 discovered DD
systems seemed massive enough to qualify as a SN\,Ia precursor. However, by
this time, theoretical simulations suggested that less than a percent of all
potentially observed white dwarfs are close DDs qualifying as SN\,Ia progenitors
\citep{1997ApJ...475..291I, 2001A+A...365..491N}, and so the failure to find a clear progenitor in 180 white dwarfs is not a surprise.

The observational focus of the 1990s on low mass white dwarfs as more
promising targets for DDs worked against the chances of finding potential
Type~Ia progenitors. Still, given the ubiquity of white dwarfs, the number of DDs found
showed that there is a large population of these binaries within
the Galaxy. Clearly a definitive test of DDs as SNe Type~Ia progenitors requires
(a) a larger sample of targets, and (b) one not selected by mass. We therefore
embarked upon a large spectroscopic survey of white dwarfs using the UVES
spectrograph at the ESO VLT UT2 (Kueyen) to search for white dwarfs and
pre-white dwarfs with variable RVs (ESO {\bf S}N\,Ia {\bf P}rogenitor
surve{\bf Y} -- SPY). There have been a number of introductory articles to SPY
\citep{2003Msngr.112...25N,2001AN....322..411N, 2005ASPC..334..375N} and many
results \citep[][to name a few]{2002A+A...386..957N,
2003A+A...400..877P,2004ASPC..318..402N,2005A+A...440.1087N,
2005A+A...432.1025K, 2007ASPC..372..387N}, but there has been no full
write-up of the original radial velocity survey. In this paper our aim is
to accomplish this for the (majority) hydrogen-atmosphere DA white dwarfs.

Since the completion of the SPY survey, there have been many discoveries of
DDs dominated by the light of ``extremely low mass'' (ELM) white dwarfs
\citep{2010ApJ...723.1072B, 2012ApJ...744..142B,2016ApJ...818..155B,2016ApJ...824...46B}, many of which will merge
well within a Hubble time \citep{2012ApJ...751..141K}. These discoveries
post-date the SPY survey since ELM white dwarfs were missed in earlier surveys
for white dwarfs due to their low gravities which makes them hard to
distinguish from early-type main-sequence stars. Some of the ELM binaries host massive white dwarfs hidden in the glare of the ELM components
\citep{2010ApJ...719.1123K, 2011ApJ...736...95M,2013ApJ...769...66B}. If the mass ratio of components in ELM DDs allows stable mass exchange, they will turn into ultracompact cataclysmic variables of the AM~CVn type
\citep{1972ApJ...175L..79F,2001A+A...368..939N}. In these systems, the
accretors can in principle very occasionally reach the Chandrasekhar mass
\citep{2005ASPC..334..387S}. Otherwise, unstable mass transfer may lead to the merger of components with formation of subdwarf~B, subdwarf~O or R~CrB stars; explosive events are not expected because of the small mass of the donors \citep{2014MNRAS.438...14D}. 

In this paper we present results of the SPY survey for the radial velocities
of white dwarfs spectral type DA (hydrogen dominated spectra). 
The project was conceived for being operationally (and scientifically) advantageous with observations in service mode at the VLT that could be executed when the telescope would otherwise have been idle \citep{1999Msngr..96...13R}. Our sample overlaps strongly
with the 615 SPY DAs analysed by \cite{2009A+A...505..441K}. \citeauthor{2009A+A...505..441K} performed a
model atmosphere analysis of SPY spectra and determined the fundamental
parameters effective temperature ($\teff$) and surface gravity ($\log g$).
Our focus is upon radial velocities, but we present additional fundamental parameter estimates to provide a uniform set of data for future comparison with theoretical models. The sample selection for SPY, observations, and data reduction are described in
Sect.~\ref{s:obs}. Our method of RV measurement is outlined in
Sect.~\ref{s:RV}. Sect.~\ref{s:results} presents the results including the
detected DDs. In Sect.~\ref{s:discussion} we discuss our results in the
context of previous surveys and implications for the evolution of DDs.

\section{Observations and data analysis}
\label{s:obs}

\subsection{Sample selection}

Targets for SPY were drawn from five sources: the WD catalogue of
\citet{1999ApJS..121....1M}, the Hamburg ESO Survey
\citep[HES;][]{2000A+A...358...77W, 2001A+A...366..898C}, the Hamburg Quasar
Survey \citep[HQS;][]{1995A+AS..111..195H, 1998A+A...338..563H}, the
Montreal-Cambridge-Tololo survey \citep[MCT;][]{2000AJ....119..241L}, and the
Edinburgh-Cape survey \citep[EC;][]{1997MNRAS.287..867K}. Our selection
criteria were spectroscopic confirmation as a white dwarf (at least from
objective prism spectra) and $B,V\le 16.5$ (depending on what information was
available) for the region south of $\delta=+15\degr$. The limit was chosen in
order to generate a sample size of order 1000 as thought necessary to uncover
a potential SN Ia progenitor
\citep[DD with $\mtot \geq \mch$,][]{2001A+A...365..491N}. We extended the survey region later to include
the region $+15\degr \le \delta \le 25\degr$, but imposed a brightness limit
of $B,V\le 15.5$. 

\begin{figure*}
\sidecaption
\includegraphics[width=12cm]{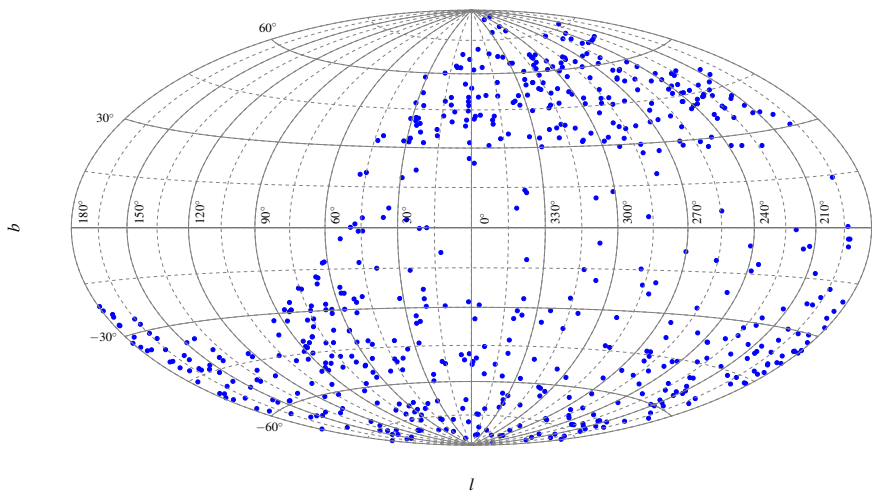}
\caption{Aitoff projection in Galactic coordinates of the DA white dwarfs observed by SPY. \label{f:ait}}
\end{figure*}

The sample of white dwarfs catalogued by \citet{1999ApJS..121....1M}
incorporates various sets plagued with different selection effects. While
early surveys for white dwarfs typically based their selection on proper
motions \citep{1979lccs.book.....L, 1978LowOB...8...89G}, later catalogues of
white dwarfs are usually a waste-product of surveys for galaxies and quasars
at high galactic latitude. One important early example is the Palomar-Green
survey \citep{1986ApJS...61..305G} which used blue colour besides the galactic
latitude as criterion. Similar approaches were used by the MCT and EC surveys
mentioned above. Since photographic prism spectra are available for the HQS
and HES surveys the presence of strong Balmer lines could be incorporated in
the selection process \citep{2001A+A...366..898C} and the colour criterion
could be somewhat relaxed.  Other white dwarfs were identified because they
showed up in EUV/X-ray surveys \citep[e.g.\ ][]{1993MNRAS.260...77P}.  While these approaches have somewhat complementary selection effects, 
there does remain a significant deficit of white dwarfs in the Galactic plane
(Fig.~\ref{f:ait}). A small selection of our own was to
avoid stars that satisfied our selection criteria but which had already been
observed by \citet{2000MNRAS.319..305M}. Three of them ended up being 
re-observed during our survey, leaving an addition of seven to the SPY 
sample\footnote{WD0401+250, 
WD0752$-$146, WD0950$-$572, WD0954+247,
WD0954$-$710, WD1407$-$475, WD2151$-$015}. We account for these stars
later when discussing the detection efficiency of the SPY sample.

We had to create a large data base of accurate coordinates and finder charts at short notice to start the service mode observations for SPY. To speed up this process we initially focussed on catalogues with good coordinates and/or
good finder charts. Important sources of the SPY observations reported in
\citet{2001A+A...378..556K} and \cite{2009A+A...505..441K} were the PG, HES,
EC, and MCT surveys, which
are all biased toward hot objects. Additional sources were the proper
motion selected white dwarfs of \citet{1978LowOB...8...89G,
  1980LowOB...8..157G} as catalogued in
\citet{1999ApJS..121....1M}. \citet{2005ASPC..334..375N} reported on results
from a ``bright'' sub-sample of SPY, which covers 93\% of all known white dwarfs 
in the survey area. This sample should be virtually free of any SPY-related selection effects. In Fig.~\ref{f:ait} we show the Aitoff projection
in galactic coordinates of the DA white dwarfs observed by the SPY project.
This figure shows the paucity of white dwarfs in the Galactic plane that we
referred to above.

\subsection{Observations}

Spectra were taken with the UV-Visual Echelle Spectrograph (UVES) of the UT2
telescope (Kueyen) of the ESO VLT. UVES is a high resolution Echelle
spectrograph, which can reach a resolution of 110,000 in the red region with a
narrow slit \citep[cf.\ ][for a description of the
  instrument]{2000SPIE.4008..534D}. Our instrument setup (Dichroic~1, central
wavelengths 3900\,\AA\ and 5640\,\AA ) uses UVES in a dichroic mode with a
$2048\times 4096$ EEV CCD windowed to $2048\times 3000$ in the blue arm, and
two CCDs, a $2048\times 4096$ EEV and a $2048\times 4096$ MIT-LL, in the red
arm. Nearly complete spectral coverage from 3200\,\AA\ to 6650\,\AA\ with only
two $\approx$80\,\AA\ wide gaps at 4580\,\AA\ and 5640\,\AA\ is achieved.  In
the standard setting used for our observations UVES is operated with an $8''$
decker in the blue arm and an $11''$ decker in the red arm. The slit was
rotated to the parallactic angle.

SPY was implemented as a service mode program. It took advantage of
observing conditions which were not usable by most other programs (moon, bad
seeing, clouds), and was run when other programs were not feasible. A wide slit
($2.1''$) was used to minimise slit losses and a $2\times 2$ binning applied
to the CCDs to reduce read out noise. Our wide slit reduced the spectral
resolution to $R=18\,500$ (0.36\,\AA\ at H$\alpha$) or better, if seeing disks
were smaller than the slit width. Depending on the brightness of the objects,
exposure times of 5\,min or 10\,min were chosen. The S/N per binned pixel (0.044, and 0.032\,\AA\ for the blue and red channel respectively) of the extracted spectrum is usually 15 or higher. Due to the
nature of the project, two spectra at different, ``random'' epochs separated
by at least one day were observed. 

Although our program was carried out during periods of less favourable
observing conditions, the seeing was often smaller than the selected slit
width of $2.1''$. This can, in principle, cause wavelength shifts, if the star
is not placed in the centre of the slit. However, since according to the
standard observing procedure the star was first centred on a narrow slit
before the exposure with the broader slit was started, it can be expected that
the star is usually relatively well centred in the slit. We used telluric
lines present in the red region of some spectra to estimate the size of the
resulting wavelength shifts (Sect.~\ref{s:telluric}) and estimated an additional
RV scatter of $0.67\mathrm{km\,s^{-1}}$.

\subsection{Data reduction}

ESO provided at the time of the SPY observations (commencing in 2000)
a data reduction pipeline for
UVES, based on MIDAS procedures. The quality of the reduced spectra is in most
cases good and we made extensive use of the pipeline reduced spectra for
selecting targets for follow-up observations and the model atmosphere analysis
of \citet{2001A+A...378..556K}.  However, sometimes the reduction pipeline
produced artefacts of varying strength, e.g.\ a quasi-periodic pattern in the
red region similar in appearance to a fringing pattern. In a few cases either
the blue or the red part of the pipeline reduced spectrum had extremely strong
artefacts of unknown origin.

Therefore we created a semi-automatic ``SPY pipeline'' to perform the final
reduction of the spectra.  Our procedure makes use of several routines of the
``UVES'' context of MIDAS, which provides reduction routines adapted for the
UVES spectrograph.  The reduction is done individually for the blue
(3200-4500\,\AA), lower red (4600-5600\,\AA), and upper red
(5700-6650\,\AA) region of the spectrum. The position of the individual
orders on the CCD is defined automatically tracing the order definition
flatfield. However, usually the star is not perfectly centred on the slit along the spatial axis in
all wavelength regions. The offset from the centre is determined and accounted
for manually, along with the width of the stellar spectrum and the
corresponding sky background area.

After accounting for cosmic ray hits and bad CCD pixels, bias and interorder
background are subtracted. Wavelength calibration is then performed
automatically using the Th-Ar reference spectrum. This is followed by the
extraction of one dimensional spectra for each order of the sky background,
flatfield, and object frames. For each spectrum, sky background subtraction and
flatfielding are performed, and finally the orders are merged. Now, the
resulting spectrum is divided by a smoothed spectrum of a DC WD,
which by definition shows no spectral features at all and therefore provides
an excellent means of correcting for the instrumental response. 
We checked a number of white dwarfs classified
DC for the absence of spectral features even at our high resolution and
selected two feature free ones (WD\,0000$-$345 and WD\,1520$-$340).

\begin{figure}
\includegraphics[bb=30 30 560 750,angle=-90,width=8.5cm]{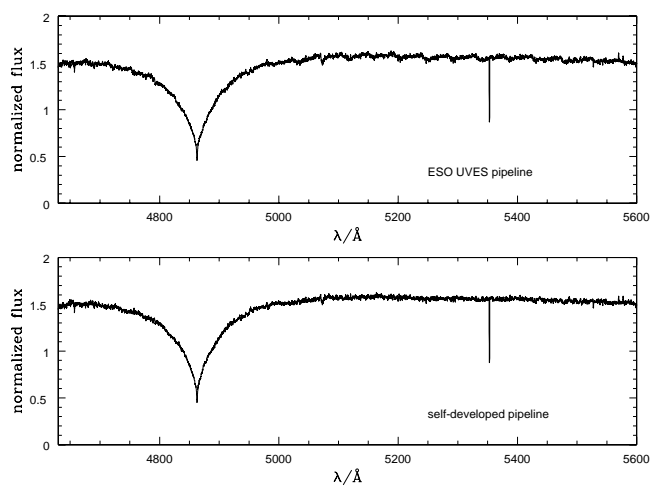}
\caption{The ripple problem and its solution. The top spectrum shows the
  reduced and merged spectrum of a DA WD. It displays a strong
  quasi-periodic pattern. The lower spectrum resulted after application of our
  ``shift correction''. } 
\label{f:ripple}
\end{figure}

A number of merged spectra did not appear smooth, but instead displayed a
quasi-periodic ripple pattern over the whole wavelength range or at least over
a significant part (Fig.~\ref{f:ripple}). The appearance is similar to the
effect mentioned above seen in some of the pipeline reduced
spectra. Obviously, this behaviour is due to a mismatch in the flux levels of
the corresponding ends of adjacent orders. Such effects should in principle be
accounted for by the flatfield division. We do not find a relation as to
\emph{when} flatfield division is performed, since the ripples remain
unchanged if we first divide the (two-dimensional) object and flatfield
frames, and then extract the individual orders. In order to investigate this
problem, we chose several stars where the signal-to-noise was large enough
that we could extract each order with the width of only one pixel, and still
obtain a useful resulting spectrum. The flatfield orders were extracted in the
same way, but with an \emph{offset perpendicular to the dispersion direction}
(hereafter \emph{separation}). We now went through a range of separations, and
found that the ripples first disappear, then appear again with different
shape. However, the separation which yields the smoothest spectrum is not
always zero and the ideal separation changes with wavelength. To
summarise, it is apparent that for each 'pixel row' of the two dimensional
object spectrum, there is only one flatfield pixel row which fits ideally, but
this can change as one moves along the order.

Given that we do not understand the problem's cause, it would seem
unreasonable to construct a complex algorithm which would take these findings
into account. However, without direct relation to the former, we found that
the reduction can be improved significantly by \emph{shifting} the flatfield
orders with respect to the object orders a few pixels in the dispersion
direction (as opposed to the perpendicular separation discussed above). This
appears counterintuitive, but it produced the desired result and the practical
implementation was easier.

The offset needed changed from night to night and sometimes between spectra
taken the same night. This was taken into account in the reduction routine,
which automatically went through a range of shift values.  We selected
spectral regions without strong features.  For each shift, the reduced
spectrum was fitted by a low-order polynomial, and a $\chi^2$ was computed
from the difference between the spectrum and the fit.  The best shift was
chosen to be the one which led to the lowest $\chi^2$ value. Note that for
each star, the shift was computed individually, although it was then applied
to the flatfield, which is the same for all stars of one night.  We emphasise
that this does {\em not correspond to a real wavelength shift of the
  spectra}. This is demonstrated by our assessment of the stability of the
wavelength scale in the Sect.~\ref{s:telluric}.

The procedure outlined above works very well for the DAs
(Fig.~\ref{f:ripple}), which are the topic of this article, and for most other
white dwarf spectral classes as well.  However, for stars of other spectral
types, e.g.\ helium-sdOs which show a large number of spectral lines, the
automatic determination of the shift does not yield a useful result because
the spectrum is too crowded with features. For these stars we determined the
best shift manually by comparing the reduction results for various shifts.
The determined shift values show no correlation with quantities like airmass
or observing time. Although usually the overall shift varies only slightly
during one night, there can be stars with completely different shift values,
which do not stand out by position, observing time, or any other quantity. We
therefore conclude that the observed effect must be intrinsic to the object
frames, not to the flatfield which is only taken once per night. However, we
could not identify an obvious reason for the origin of this problem.

We also included the observations from the later run 072.D-0487(A), taken
during January, February and March 2004. This last set of data was reduced
using the Reflex pipeline \citep{2013A+A...559A..96F}.  Reflex is an
environment that was developed to provide an easy and flexible way to reduce 
VLT/VLTI science
data using the ESO pipelines. The output data after using the pipeline are
wavelength and flux calibrated spectra. Note that the ripple problem occurs less often in the latest pipelines.

\subsection{Telluric lines}
\label{s:telluric}

A number of telluric absorption features are known in the red spectral
range. In the \halpha region these are mostly caused by water in the
atmosphere. Overall Paranal is a very dry place and thus the water features in
our UVES spectra are usually weak.  However, since the SPY spectra were taken
preferentially during periods of relatively poor weather conditions, stronger
telluric absorption is present in a number of spectra. Since the RV
determination can be skewed if a strong telluric feature coincides with the
stellar H$\alpha$ core, we corrected the telluric contribution.

\begin{figure}
\includegraphics[width=6.6cm,angle=-90]{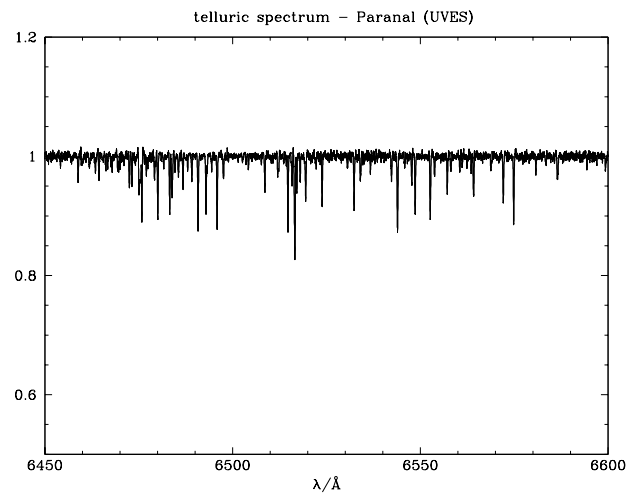}
\caption{Telluric template spectrum in the H$\alpha$ region.
\label{f:tellurium} 
}
\end{figure}

For this purpose we produced a telluric UVES template, which was created from
50 UVES spectra selected from the complete SPY data set of $\approx$2000
spectra on the basis of strong water absorption and simple spectral appearance
in the red range. This included white dwarfs of spectral types DA, DB, DBA,
and DC. For each spectrum the local continuum was evaluated in an iterative
procedure. The spectra were convolved with a Gaussian of 1\,\AA\ FWHM. Data
points more than $3\times$ the noise level below the continuum were
clipped. This procedure was repeated until the number of clipped wavelength
points remained constant.  The noise level was estimated from neighbouring
featureless parts of the spectra. The inner line cores of \halpha, if present,
were excluded by hand. Since the white dwarf sample covers a large range of
RVs, this does not produce gaps in the final co-added spectrum.  Finally each
spectrum was divided by the smoothed continuum. The individual spectra were
rebinned to a common wavelength scale and the absorption was scaled to a
``normalised telluric absorption strength''. Afterwards all 50 individual
telluric spectra were co-added to the template spectrum shown in
Fig.~\ref{f:tellurium}.

To correct the telluric contamination of the stellar spectra, we measured the
strength of the telluric features in the 6450\,\AA\ to
6500\,\AA\ interval. No sharp stellar lines are present in this range.  The
template was scaled accordingly and the correction was done by dividing the
stellar spectrum by the scaled template.
 
The presence of telluric lines complicates the analysis of the stellar
spectra. However, there is a bright side as well. Telluric absorption lines
can be used to measure RV velocities to very high precision
\citep{1973MNRAS.162..243G}. We checked the RV stability of the UVES/SPY
observations by cross-correlating the template with individual spectra with
strong telluric features. The resulting scatter is equivalent to $\pm
0.67\,\kms$ indicating good stability; this number was added in quadrature to
the radial velocity uncertainties described in the next section when it came
to judging evidence for radial velocity variability.

\section{Radial velocity measurement}
\label{s:RV}

The radial velocities (RVs) of DA can be most accurately measured from the
sharp line cores of the Balmer lines. These are most pronounced in \halpha,
and to an lesser extent in \hbeta (Fig.~\ref{f:sample}). These line cores are
formed in high layers of the stellar atmosphere, where the density is low and
pressure broadening small\footnote{\citet{2015ApJ...808..131H} carried out
  theoretical calculations and claimed that pressure shift can mimic RV shifts
  of tenth of \kms, if the inner line wings of \hbeta are measured. The
  authors found negligible effects for \halpha. Our empirical test presented
  in Appendix~\ref{s:pressure} indicates a much smaller effect, if present at
  all.}.  The low densities allow the occupation numbers to deviate from local
thermal equilibrium (LTE) and allow for NLTE effects, which are responsible
for the strength of the cores \citep{1973A+A....25...29G}. Since most white dwarfs are
slow rotators \citep{1988ApJ...332..910K, 1997A+A...323..819H,
  1998A+A...338..612K}, the line cores are basically Gaussian with widths
corresponding to the temperature in the formation layer.

Radial velocities of stars are often measured using cross-correlation
algorithms. Observed spectra are correlated against model spectra or observed
template spectra. The advantages of this method are the flexibility and its
straightforward usage. As discussed in \citet{2001AN....322..411N}, the broad
line wings of the Balmer lines cause complications if the traditional
cross-correlation method is applied to white dwarfs. Thus we developed a
modification of the cross-correlation method, which is based on a $\chi^2$
test \citep{2001AN....322..411N} and measured the relative RV shifts between
the observed spectra. This proved to be a quick and flexible way to select RV
variable stars for follow-up observations
\citep[e.g.\ ][]{2001A+A...378L..17N}. However, experience showed that this
method has its drawbacks (error estimates are not always reliable and
sometimes spurious results are produced, especially for stars with weak line
cores), and thus we developed more sophisticated methods for the final
analysis.  Another motivation for a different approach to the RV measurements
was our interest to use the absolute RVs for a study of the space velocities
of the white dwarfs in the SPY sample \citep{2003A+A...400..877P,2006A+A...447..173P,2007ASPC..372..107R}.

Several investigations \citep{1988ApJ...334..947S, 1991ApJ...374..281F,
  1995MNRAS.275..828M, 1999MNRAS.307..122M} modelled the line profiles with
Lorentzians or Gaussians. However, these functions describe the line profiles
only approximately and combinations of three or four functions can be necessary
for an accurate modelling \citep{1999MNRAS.307..122M}. Model atmosphere line
profiles on the other hand can reproduce the observed Balmer lines of DA white dwarfs
 quite accurately \citep{1997A+A...323..819H}, if the NLTE effects are
taken into account. \citet{2001A+A...378..556K} determined fundamental
parameters of about 200 white dwarfs observed during the first nine months of
the SPY observations, 163 of them of spectral type DA (compiled in their
Table~A.1). \citet{2009A+A...505..441K} obtained the parameters of a much larger sample (including 615 DA white dwarfs). Thus, all the necessary information to apply line profile fits is available for our sample.

\subsection{Model atmospheres}

%Depending on the temperature we used two grids of model spectra for the RV
%determination.  In both cases the atmosphere were computed for a pure
%hydrogen composition, which is appropriate for most DA white dwarfs.  Above
%20\,000\,K the self-consistent NLTE model atmosphere of
%\citet{1999ApJ...517..399N} were applied. Below these temperatures convection
%and additional opacity by quasi-molecular satellites of the Lyman lines
%become important. These effects were not included in the
%\citet{1999ApJ...517..399N} calculations.

Radial velocities are derived by fitting the profiles of
the \halpha and \hbeta Balmer lines with model spectra and computing the
optimum RV. In the few cases with a clear detection of photospheric
Ca~II H\&K lines, these were also included.
White dwarfs with
temperatures above 20\,000\,K were analysed with the grid of self consistent
pure hydrogen NLTE model atmospheres for DA white dwarfs of
\cite{1999ApJ...517..399N}. Cooler white dwarfs were analysed with the grid
described in \citet{1998A+A...338..612K}. These model spectra used NLTE 
line formation physics as applied to the LTE atmosphere models of D.~Koester.

Fundamental parameters are derived from
the spectral analysis of the complete Balmer series as described in
\citet{2009A+A...505..441K} for the SPY DA sample. The input physics of
the model grids used for determining the RVs is not exactly the same as for the
models used for the \citet{2009A+A...505..441K} analysis. Thus we redetermined
the fundamental parameters temperature and gravity with the grids actually
used for the RV measurement and the fitting programme \fitsb described
below. This ensures that the fitting process is self-consistent. 

In the case of double-lined systems, temperatures and gravities of both
components were determined. This was an iterative process with a first fit
carried out using first guess RVs. The estimated fundamental parameters were
then used for determining RVs from fitting the \halpha and \hbeta lines. These
initial estimates were refined by refitting temperature/gravity and RVs until
parameter changes in subsequent steps became negligible. Model spectra for the
two parameter sets were coadded after scaling them accordingly using the
mass-radius relations derived from the tracks of \citet{1999MNRAS.303...30B}
complemented by the \citet{1961ApJ...134..683H} zero temperature models for
the highest masses. During these fits, the white dwarf mass-radius relation was used as a constraint to reduce parameter degeneracy.

\subsection{FITSB2 and the fit process}
\label{s:fit_process}

The programme \fitsb can fit line profiles and measure RVs. It was originally
developed to analyse the spectra of double-lined (SB2) systems, but can be
applied to spectra of single-lined systems (RV variable or not) as well.

The basic philosophy for fitting line profiles follows the {\sc fitprof}
program described in \citet{1999ApJ...517..399N}. Both, the observed and
theoretical Balmer line profiles are normalised to the local continuum in a
consistent manner. Wavelength shifts are calculated for the actual RV estimate
for each spectrum. The synthetic spectra are convolved to the observational
resolution with a Gaussian and interpolated to the actual parameters with
bi-cubic splines and interpolated to the observed wavelength scale.

Parameters are determined by minimising $\chi^2$. The optimisation is done by
means of a simplex algorithm, which is based on the {\sc amoeba} routine of
\citet{1992nrfa.book.....P}. The advantages of this approach are a high degree
of stability and flexibility. New functions and parameters can be easily
integrated. One disadvantage is that the convergence of simplex algorithms is
not very fast and thus CPU time intensive. However, in practice this is not a
major limitation for line profile fitting with \fitsb.

\fitsb offers a number of options for fitting line profiles: model spectra,
Gaussians, Lorentzians, and polynomials or a combination of these four. Most
white dwarf spectra were simply fitted using model profiles calculated for the
appropriate parameters. The default fit interval was $\pm$15\,\AA\ relative to
the line centre. However, sometimes it was appropriate to use larger
wavelength intervals. The intervals were adjusted according to the measured
RVs. In a number of cases the model line profiles did not give a
satisfactory fit of the observed spectra, potentially compromising the
accuracy of the measured RV. If that happened Gaussians, Lorentzians and/or
polynomials were used to improve the fit; see Sect.~{\ref{s:results}
for a discussion and examples.

Our observational set-up results in vastly oversampled spectra. Moreover,
Doppler broadening results in a substantial broadening of the NLTE line cores
of the \halpha and \hbeta (the Doppler width of \halpha at 10\,000\,K is
2.2\AA). This allowed us to use a simple method to estimate the local noise. A
moving parabola was fitted to 20 pixels %(corresponding to 0.8\,\AA)
and the
standard deviation computed. This was repeated for all pixels. Afterwards, the
pixel by pixel standard deviations were smoothed by a 50 pixel boxcar.

Radial velocity errors were calculated using bootstrapping. This method
randomly selects $N$ points of the $N$ points of the observed spectra. Points
can be selected more than once, which has the effect that on average
$1-e^{-1}$ points of the original spectrum are represented in the new
spectrum. Radial velocities are refitted using the bootstrapped spectrum. This
process is repeated and the errors are computed from the standard deviation of
the derived parameters. Error estimates using bootstrapping are better at taking into account uncertainties resulting from
imperfections of the input data (e.g.\ hot or dead pixels) and non-Gaussian
noise than standard estimates from the co-variance matrix.

Error estimates from bootstrapping are subject to statistical fluctuations and
a sufficient number of bootstraps need to be done.  The bias-corrected three
step method of \citet{2000Eco..68...23} is applied to estimate the number of
necessary iterations. We required that the probability of the error estimates
being wrong by more than 10\% is less than 1\%. The number of bootstraps
depends on the distribution of bootstrapped parameters with long tails
requiring more bootstrap steps to achieve the required accuracy. Our
accuracy requirements resulted in a minimum of 332 bootstraps. Note that the
parameters describing Gaussians, Lorentzians and polynomials modifying the
model spectrum are treated as free parameters and thus enter the error budget.

\begin{figure*}
\includegraphics[height=17.2cm,angle=-90]{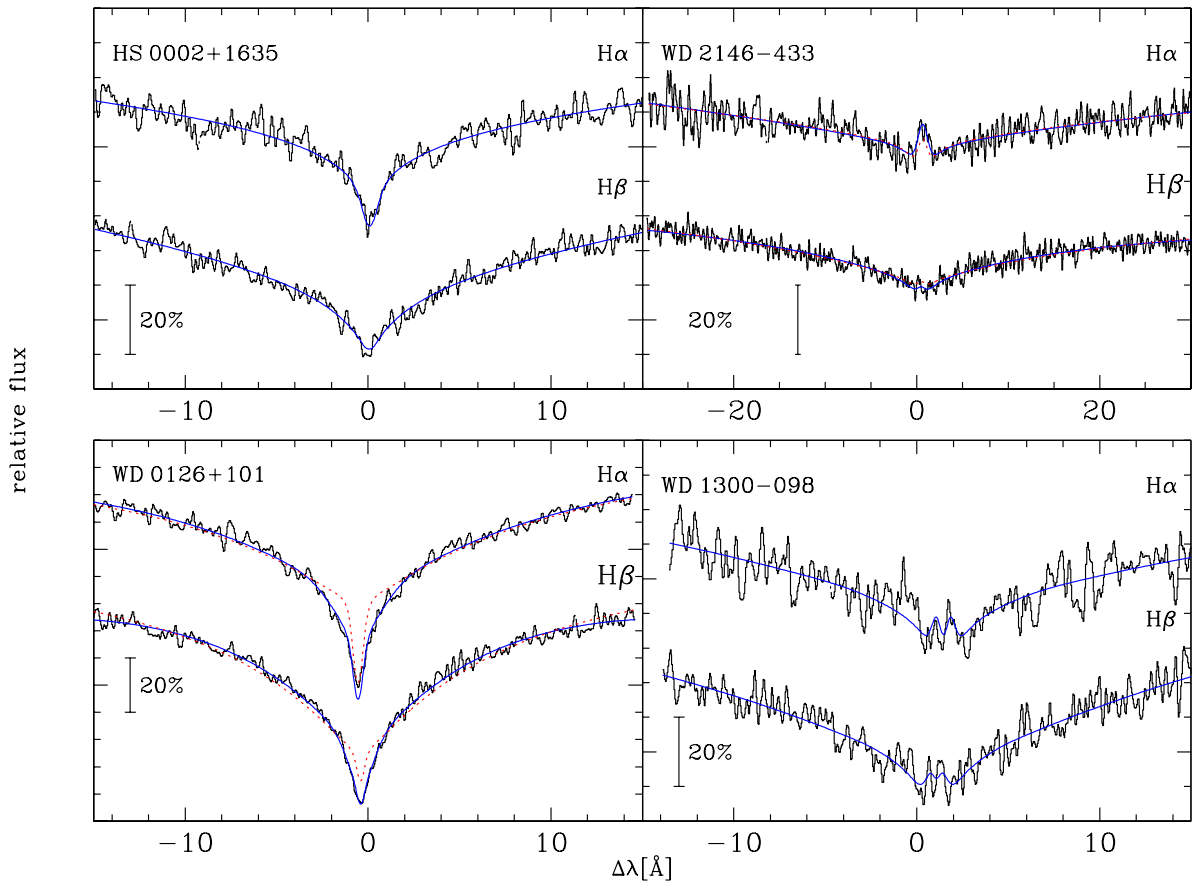}

\caption{Observed H$\alpha$ and H$\beta$ profiles compared to the fitted model
  spectrum. The graph shows one observed spectrum for each binary and the best
  fitting model spectrum (blue smooth line). Here and in the following figures
  we display the observed spectra smoothed with a Gaussian of
  0.15\,\AA\ FWHM, hardly affecting the spectral resolution. 
  $\Delta \lambda$ is the wavelength offset relative to the
  laboratory wavelengths of \halpha and \hbeta, respectively.  
  {\it Top left:} HS\,0002+1635, fitted with model spectra without 
  modifications. {\it Top right:} WD\,2146$-$433, a hot DA with a Gaussian 
  added to better reproduce the central emission core (model spectrum without 
  correction shown as dashed red line.)
  {\it Bottom left:} WD\,0126+101, one of the cool DAs displaying a broader 
  and stronger central core than predicted by the models (model spectrum 
  without correction shown as dashed red line.).  
  {\it Bottom right:} WD\,1300$-$098, one of the magnetic white dwarfs.
\label{f:sample}
}
\end{figure*}

\section{Results}
\label{s:results}

Survey spectra were taken for a total of 689 targets classified as DA. It
turned out that the spectra of 46 stars revealed the presence of red
companions \citep{2009A+A...505..441K}, almost all M type main sequence stars,
but also at least one brown dwarf \citep{2006Natur.442..543M}. The DA+dM
nature of most of these targets was not known at the time they were scheduled
for observations. For many of the white dwarfs only low quality classification spectra
of the blue range existed, which did not allow the detection of the cool
companion. 

A further 18 targets could not be checked for RV variations, because only one spectrum was taken, leaving 625 white dwarfs we could analyse for RV variations and the presence of a compact companion.  Our RV measurements of the 643 DAs without a cool companion are summarised in Table~\ref{t:RV}. We applied heliocentric corrections to the velocities, but no corrections for gravitational redshift. The latter can exceed 100\,\kms\ for high mass white dwarfs and need to be taken into account when calculating space velocities, but in this study the emphasis is on velocity variations.

We used the same criterion for variability as used by
\citet{2000MNRAS.319..305M}. For each star a weighted mean of the RVs is
computed. We then calculate $\chi^2$ from the deviations of the individual RVs
from the mean. $\chi^2$ measures the goodness of fit of the ``model
assumption'' of a constant RV. The comparison with a $\chi^2$ distribution
with the appropriate number of degrees of freedom yields the probability $p$
of obtaining the observed value of $\chi^2$ or higher from random
fluctuations.  The computed values of $\chi^2$ and $\log p$ are tabulated in
Table~\ref{t:RV} and the distribution of $\log p$ is displayed in Fig. \ref{f:logp}.  We chose a detection threshold of $p<10^{-4}$, i.e.\ white dwarfs
with $p$ values below this value are considered to be binaries (indicated as
``DD'') in Table~\ref{t:RV}. This limit corresponds to 0.1 expected false
detections due to random chance in the complete SPY sample of
$\approx$1000 white dwarfs and pre-white dwarfs. Two targets, WD0032-317 and WD1241-010, have $p < 10^{-3}$, indicative of a higher-than-average chance of their being binaries, but do not meet our $p < 10^{-4}$ criterion. WD1241-010 is in fact a known DD \citep{1995MNRAS.275..828M}; WD0032-317 could reward future study. Three targets did not show RV variations
exceeding our detection threshold, but visual inspection identified them as double-lined systems, consisting of two white dwarfs. They are indicated by
``dd'' in Table~\ref{t:RV}. See Sect.~\ref{s:DD} for a discussion of individual objects identified this way.

\begin{figure}
\centering
\includegraphics[width=\columnwidth]{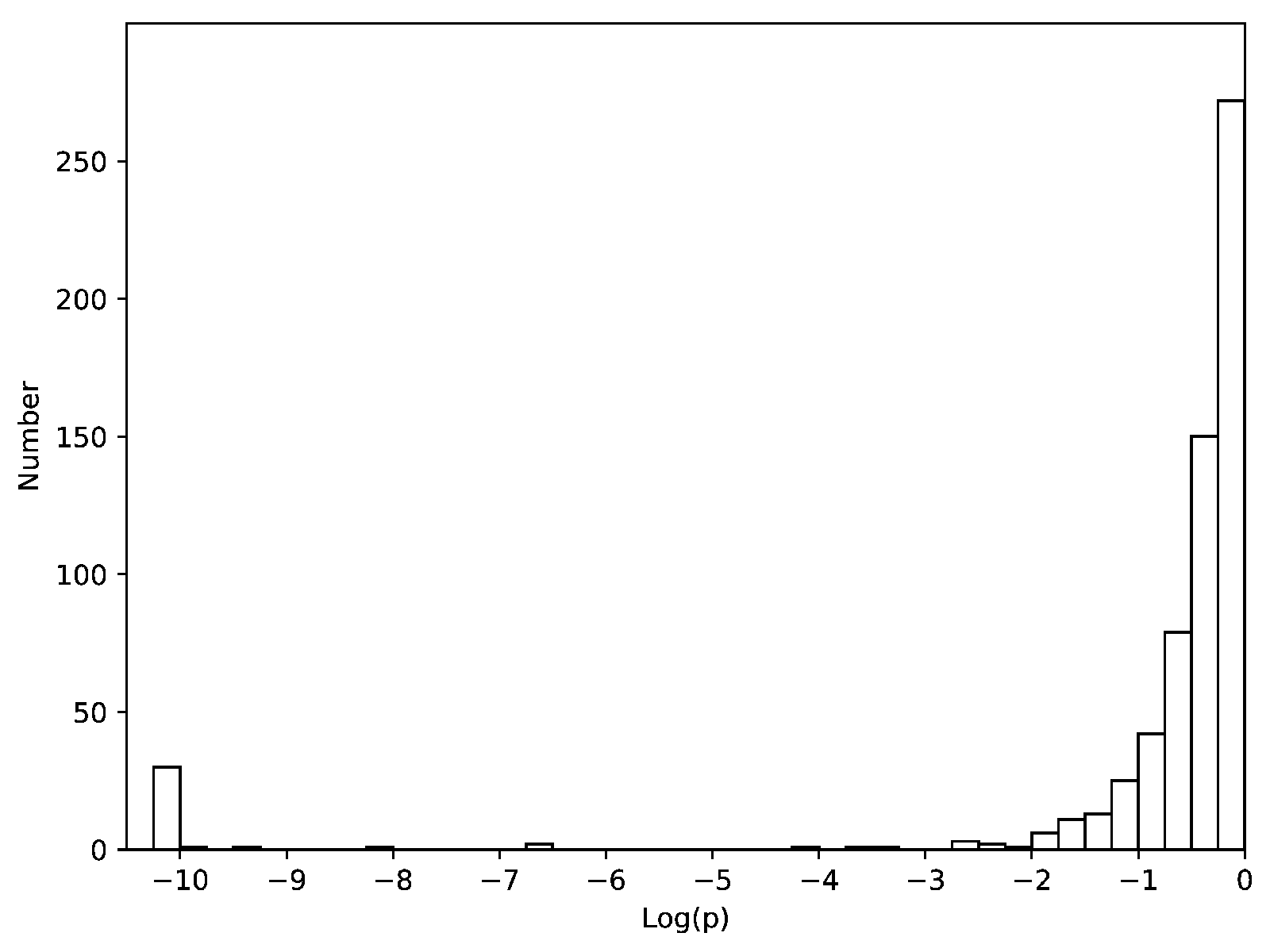}
\caption{Distribution of $\log p$ from Table~\ref{t:RV}. The peak on the right represents statistical noise only. Our detection threshold is $\log p < -4$. Two systems have $-4 < \log p < -3$;
the bin at $\log p =-10$ summarises all objects with $\log p \le -10$; many of them have $\log p$ much less than this value.
\label{f:logp}}
\end{figure}

When a combination of model
atmosphere spectra with Gaussians, Lorentzians, and/or polynomials is used for
the line profiles fitting this is marked in the comments column of 
Table~\ref{t:RV}. See Sect.~\ref{s:RVtable} for a detailed description of the 
table.

\subsection{Peculiarities}
\label{s:pec}

As stated above, the fits of the \halpha and \hbeta line make use of model
atmosphere profiles. These are computed for the fundamental parameters derived
from the fit of the entire Balmer line series as outlined in
\citet{2009A+A...505..441K}. Thus in an ideal world the only free parameters
during the fitting procedure would be the RVs. This worked well for about half
of the programme stars (e.g.\ for HS\,0002+1635 in the top left panel of
Fig.~\ref{f:sample}). However, the fit between observed and synthetic line
profiles was less than perfect for very hot and for cool white dwarfs.

\paragraph{Very hot white dwarfs.} The hottest stars in the SPY sample displayed 
stronger emission cores than predicted by the models (example WD\,2146$-$433;
Fig.~\ref{f:sample}, top right). The problems are probably explained by
the presence of metals in the atmosphere, which modify the atmospheric
structure and thus the line profiles \citep{1996ApJ...457L..39W}. However,
\citet{2015A+A...579A..39L} report that discrepancies still persist, if the
correct abundances are adopted for model calculations and a good fit can only
be achieved with arbitrarily increased abundances. We took the pragmatic
approach of amending the cores of our model spectra by adding Gaussians and/or
Lorentzians to achieve a good fit.

\paragraph{Cool white dwarfs.} Many cool DAs showed broader and stronger line cores 
than predicted by the models (e.g.\ WD\,0126+101; Fig.~\ref{f:sample}, bottom
left).  \citet{1998A+A...338..612K} reported on the presence of flat bottom
profiles of the cores of \halpha for pulsating white dwarfs of the ZZ~Ceti
type. The reason for these flat cores is very likely the velocity field
associated with the pulsations \citep{2007A+A...473..239K}.  Similar and even
stronger effects are present in our sample for a number of cool white dwarfs
outside the ZZ~Ceti instability strip \citep[WD\,0126+101 has a temperature of
  8557\,K;][]{2009A+A...505..441K}. One can speculate that the reason of this
discrepancy are violent convective motions in the atmospheres as seen in
hydrodynamic models \citep{2013A+A...552A..13T}.

A more detailed comparison with models could give interesting insights, but
here we took again a pragmatic approach and combined the model atmosphere spectra
with Gaussians, Lorentzians, and/or polynomials to achieve a good fit of the
line cores.

\paragraph{Magnetic white dwarfs.} The SPY sample includes ten magnetic white dwarfs, 
most of them new discoveries based on the Zeeman splitting of the Balmer lines
\citep{2009A+A...505..441K}. Our model spectra do not include the effects of
magnetic fields. We simply modelled the observed spectra as well as possible
using combinations of Gaussians and Lorentzians (see WD1300$-$098;
Fig.~\ref{f:sample}, bottom right). This is a rather crude and unphysical
approach, but once a good reproduction of the line profiles is achieved,
accurate RV velocities can be measured. Note that the absolute velocities
measured this way, should be treated with caution. For that the measurements
should be repeated with a proper treatment of the magnetic fields. None of the
magnetic white dwarfs displayed any RV variation.

\paragraph{Spectral types DAB and DAO.}

Four of the white dwarfs in our sample are of spectral type DAB and two of spectral
type DAO (HE\,1106$-$0942\footnote{This star is amongst the hottest and is the most luminous star in our sample.  \citet{2007A+A...462..269S} classified it as a subluminous O star (sdO) and derived atmospheric parameters similar to those listed in Table \ref{t:tefflogg}.} and WD\,1305$-$017) showing lines of He~I or He~II, respectively in
addition to the hydrogen lines. We used NLTE atmospheres computed for a mix of
hydrogen and helium for analysing the DAOs. 

HS\,0209+0832 is a genuine DAB white dwarf showing helium line profile variations \citep{1997A+A...324L..53H}.
Two other DABs (WD\,0128$-$387 and WD\,0453$-$295) were analysed by
\citet{1994ApJ...429..369W}, who showed that a consistent fit of lines and
spectral energy distribution is only achievable with a composite spectrum
combining a hydrogen-rich and a helium-rich WD. Similarly, \citet{2002ApJ...566.1091B} found WD\,1115+116 to be a DA+DB double-lined binary system from a model atmosphere analysis. This was confirmed by \citet{2002MNRAS.334..833M}'s measurement of a 30~day orbital period, the longest measured for any DD. A detailed decomposition of the spectrum is beyond the aims of this investigation and the He~I
lines are weak. We fitted the hydrogen-rich component with a DA model
atmosphere and used Gaussians to fit He~I lines. We included only the
4713.3\,\AA, 5015.7\,\AA\ and 5875.8\,\AA\ lines not displaying strong
forbidden components. Results for the DABs will be discussed in
Sect.~\ref{s:DD}. 

\begin{figure}
\centering
\includegraphics[width=9cm]{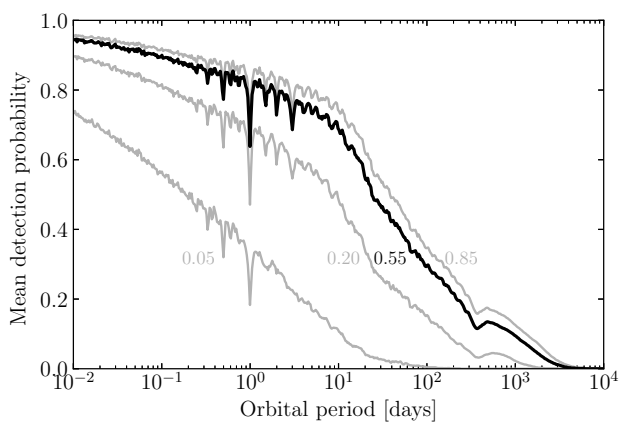}
\caption{The mean detection efficiency of the 632 targets with more than one
RV in our sample as a function of orbital period, including 7 additional 
targets from \protect\citet{2000MNRAS.319..305M} that satisfy the SPY survey criteria. Dips in efficiency at e.g.\ $P=1$\,d correspond to typical time intervals between exposures. The calculations account for randomly inclined orbits and orbital phases and were based on the assumption that the visible white dwarf has a mass of $0.55\,\msun$ and companion masses of $0.05$ (brown dwarf), $0.20$ (ELM), $0.55$ (typical white dwarf) and $0.85\,\msun$ (Chandrasekhar mass system) as indicated on the plot.\label{f:detection}}
\end{figure}

\subsection{Detection efficiency}

The mean detection efficiency of the SPY sample is plotted in
Fig.~\ref{f:detection}. For each star observed, the detection probability 
was evaluated as a function of orbital period using the same times and radial velocity uncertainties as observed along with the $0.67\,\kms$ uncertainty
added in quadrature that we discussed earlier. The probabilities were obtained by calculating the critical orbital inclination $i_c$ for which our $\log p = -4$ threshold was met, and then using $P(i \ge i_c) = \cos(i_c)$ on the assumption of
randomly inclined orbits. Random orbital phase offsets were accounted for by 
averaging over a uniform grid of 20 initial orbital phases in each case. We included the radial velocities measured by \cite{2000MNRAS.319..305M} for 10 targets that fulfil the SPY criteria but which we had excluded from the survey
because of the earlier study. Three of these targets were in fact re-observed in SPY and so the combined datasets were used for these. The remaining seven were a small addition to  overall sample, to give a total of 632 targets.

The detection probabililty of a given DD depends upon the  masses of the component stars. We assumed a mass of 
$M_1 = 0.55\,\msun$ for the dominant primary star, while the assumed companion mass $M_2$ differentiates
the different lines plotted in Fig.~\ref{f:detection}; the lines shown are invariant for constant values of
$M_2^3/(M_1+M_2)^2$. The key point of the plot is that SPY has good sensitivity over all of the known range of DD orbital periods. For typical orbital periods of DDs of around a day and normal mass white dwarf companions, the detection probability is of order 80\%, while it exceeds 90\% for periods of a few hours or less. The high precision of the SPY velocities means that there is even significant sensitivity to orbital periods as long as 100 days or more. Low mass companions cause lower amplitude RV changes, and so the detection probability drops significantly for brown dwarf mass companions ($M_2 \sim 0.05\,\msun$), but, since all known examples of unresolved white dwarf/brown dwarf binaries are of short period \citep{2017MNRAS.471..976P}, $\sim 10\,$hours or less, SPY has significant  sensitivity to such systems as well.
Calculations such as those of Fig.~\ref{f:detection} could be used to correct for DDs missed in SPY \citep[WD1241-010 is one such system;][]{1995MNRAS.275..828M}, but we refrain from doing this as the correct approach is a more nuanced multi-dimensional problem that allows for multiple selection effects when comparing theoretical and observational populations, which we discuss in more detail in Section~\ref{s:selection}.

\subsection{DA white dwarfs with Ca~II lines}

A number of DA white dwarfs observed by SPY display Ca~II absorption
lines \citep{2005A+A...432.1025K} and are thus of spectral type DAZ.  The most
prominent features are the Ca~II H and K resonance lines.  These lines
are much narrower than the \halpha and \hbeta Balmer lines used by us and the Ca\,K line is well suited for accurate RV measurements. The other
component of the doublet, Ca\,H, is usually too strongly disturbed by the
near-by strong \heps line to be useful.

The Ca~II lines are broadened by quadratic Stark (and eventually
neutral) broadening, which both lead to small asymmetries and line
shifts. This could be a potential problem in particular, when lines are very
strong, but if $\chi^2$ is small then we can be confident that the impact is
too small to make a noticeable effect.

%Apparently this is too small to make a noticeable effect, but it should be
%tested and mentioned. HAS THIS BEEN TESTED?

Nine of the DA white dwarfs of the sample presented show detectable
Ca~II lines of photospheric origin. This was established by
\citet{2005A+A...432.1025K} from a comparison of the RV measured from the
calcium lines and the Balmer lines. Interstellar Ca lines are present in the
spectra of some hotter white dwarfs.

\begin{table}
\caption{Radial velocities $v(\mathrm{Ca})$ measured from photospheric Ca\,K
  lines compared to the Balmer line values $v(\mathrm{H})$.}
\label{t:CaRV}
%\begin{tabular}{ll@{\hspace{0.5ex}}lr@{$\pm$}lr@{$\pm$}l}\hline\hline
\begin{tabular}{lllr@{$\pm$}lr@{$\pm$}l}\hline\hline
object &date &UT &\multicolumn{2}{c}{$v(\mathrm{Ca})$} 
       &\multicolumn{2}{c}{$v(\mathrm{H})$}\\ 
       &     &     &\multicolumn{2}{c}{\kms} 
       &\multicolumn{2}{c}{\kms}\\ \hline
WD\,0408$-$041    &15/09/00 &09:28 &19.2&1.5   &19.8&1.3\\
                  &02/09/01 &09:17 &17.4&2.0   &18.7&1.7\\
WD\,1124$-$293    &23/04/00 &03:55 &29.4&1.5   &29.5&0.7\\
                  &17/05/00 &02:07 &29.3&2.2   &29.2&1.5\\
WD\,1150$-$153    &23/04/00 &04:27 &23.7&1.1   &28.4&1.5\\
                  &19/05/00 &01:23 &19.3&0.9   &22.6&1.4\\
WD\,1202$-$232    &23/04/00 &04:44 &22.6&2.3   &23.1&0.3\\
                  &17/05/00 &03:08 &25.1&2.3   &23.6&0.4\\
WD\,1204$-$136    &23/04/00 &04:54 &37.5&1.1   &37.5&1.4\\
                  &17/05/00 &03:31 &35.3&1.7   &38.1&3.2\\
HE\,1225$-$0038   &22/04/00 &02:44 &11.6&2.4   &10.6&0.7\\
                  &17/05/00 &03:44 &10.7&3.0   &12.1&1.4\\
                  &04/07/00 &23:28 &13.2&3.4   &11.1&0.6\\
HE\,1315$-$1105   &23/04/00 &01:30 &32.1&2.5   &31.9&1.0\\
                  &19/05/00 &03:08 &31.4&2.3   &31.5&1.3\\
WD\,1457$-$086    &13/07/00 &04:07 &20.5&3.6   &22.9&2.0\\
                  &16/07/00 &02:03 &21.4&3.0   &19.4&1.6\\
WD\,2326+049      &06/08/00 &04:18 &43.4&0.7   &44.4&1.1\\ 
                  &17/09/00 &04:02 &39.3&0.5   &39.8&1.2\\ \hline
\end{tabular}
\end{table}

\begin{figure}
\includegraphics[width=6.6cm,angle=-90]{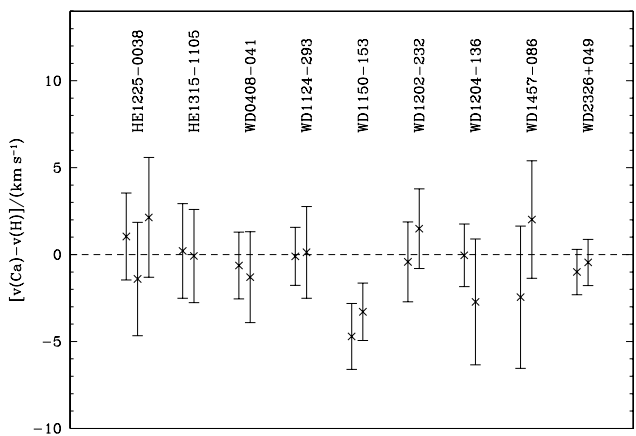}
\caption{Differences between $v(\mathrm{Ca})$ and $v(\mathrm{H})$ for
  the individual spectra as listed in Table~\ref{t:CaRV}. Error limits
  are computed by combining the error estimates for the lines in quadrature.
\label{f:CaRV}}
\end{figure}

Here we will use the photospheric Ca lines to check the internal
consistency of the RV measurements.  While the \halpha and \hbeta lines are
registered with the two CCDs in the red channel of UVES, the Ca\,K lines are
observed with the third CCD in the blue channel. CCD specific problems or
systematic errors during the spectral reduction process and RV measurement
should show up in a comparison.

Results for the Ca\,K line and the corresponding Balmer line measurements are
listed in Table~\ref{t:CaRV} and a plot of the resulting differences is shown
in Fig.~\ref{f:CaRV}. The agreement is very good. Only two values out of the
19 in Fig.~\ref{f:CaRV} show deviations exceeding the $1\sigma$ limits
computed from the individual error limits. The resulting reduced $\chi^2$
amounts to 0.72, well in agreement with purely statistical scatter. We
conclude that SPY was routinely able to measure radial velocities of white dwarfs 
at the few $\kms$ level, and, on brighter targets, to well below
$1\,\kms$.

%{\bf Telluric lines do not give systematic problems blue and red gives the
%same.  If you use two contimuum points, you get a discrepancy, while if you
%use the line method to normalize then it is fine. The formal erorr are
%smaller if you use two continuum points.  }

\subsection{Double degenerates}
\label{s:DD}

Our RV measurements are summarised in Table~\ref{t:RV}. RV variable DDs are
indicated by ``DD''. We detected a total of 39 DDs (see Table~\ref{t:RVDD} for
more details). Of these, 19 are single-lined (SB1) and 20 double-lined
(SB2). For ten of the SB2 systems radial velocity curves have been measured and masses of the components derived. Ten SB2 systems are still awaiting follow-up spectroscopy. Similarly, for nine SB1 systems radial velocity curves have been measured and minimum masses of the components derived. Four of these DDs have been published in previous works not related
to SPY, while three others have orbits presented in papers unrelated to SPY but subsequent to their detection as DDs in SPY -- see Table~\ref{t:RVDD} for references.

\begin{table*}
\begin{center}
\caption{List of the DDs identified in this survey. References
are provided for those that have previously been identified as DDs,
in many cases from the SPY survey. The mass of the primary $M_1$ is given
(see appendix \ref{s:fundamental}). $M_2$ and $P$ are secondary masses and periods
determined by follow-up observations. \label{t:RVDD}}
\begin{tabular}{lcclllp{8cm}}
\hline
\hline
\noalign{\smallskip}  
Object & $\log p$ &Type &$M_1$ &$M_2$ &P & References \\
       &           &     &($M_\odot$) &($M_\odot$)  &(d) &\\
\noalign{\smallskip}
\hline
%%\input{RV_AA_DD_postref.tab}
%%
%% TRM 17/04/2019: inserted contents of above table directly into
%% tex file because I couldn't edit it on overleaf somewhat surprisingly
%% this to allow me to change content so the RV_AA_DD_postref.tab file 
%% is out of date
WD0028$-$474 		&$<-$100&SB2 &0.54 &$\ge 0.46$ &0.39 & SPY: {\cite{2009A+A...505..441K}}, {\cite{2007ASPC..372..387N}}. {\cite{2017MNRAS.466.1575R}}\\
 WD0037$-$006 		&$<-$100&SB2 &0.51 & && \\
 WD0101+048 		&$-$17.54 &SB1 &0.49 & &&\\
 WD0128$-$387           &$-$0.48&SB2   &0.85 & && \\
 HE0131+0149 		& $-$9.47 &SB1 &0.50 & &&\\
 WD0135$-$052 		&$<-$100&SB2 &  &  & 1.56 & Mass could not be derived (see Sect.~\ref{s:fundamental}); {\cite{1988ApJ...334..947S}}, {\cite{1989ApJ...345L..91B}}\\
 MCT0136$-$2010           &$-$0.39  &SB2 &0.86 & &&\\
 HE0205$-$2945 		& $-$4.07 &SB2 &0.41 & &&\\
 WD0216+143 		& $-$6.75 &SB1 &0.54 & &&\\
 HE0225$-$1912 		&$-$28.02 &SB2 &0.55 &0.23 &0.22& SPY: {\cite{2007ASPC..372..387N}} \\
 HS0237+1034            &0.00     &SB2 &0.67 & &&\\
 HE0315$-$0118 		&$<-$100 &SB2 &0.50 &0.49 &1.91&SPY:
 {\cite{2009A+A...505..441K, 2007ASPC..372..387N}}. {\cite{2017MNRAS.466.1575R}}\\
 HE0320$-$1917 		&$-$42.87 &SB1 &0.31 &$\ge 0.45$ &0.86 &SPY: {\cite{2005A+A...440.1087N}}, {\cite{2009A+A...505..441K}} \\
 HE0324$-$1942 		& $-$8.18 &SB2 &0.78 & &&\\
 HE0325$-$4033 		& $-$9.86 &SB1 &0.49 & &&\\
 WD0326$-$273 		&$<-$100 &SB1 &0.36 &$\ge 0.96$ &1.88&SPY: {\cite{2005A+A...440.1087N}}, {\cite{2009A+A...505..441K}}  \\
 WD0341+021 		&$<-$100&SB1 &0.38 & && Probably long period, {\cite{2000MNRAS.319..305M}}\\
 WD0344+073 		&$-$19.26 &SB1 &0.39 & && \\
 HE0410$-$1137 		&$-$40.63 &SB2 &0.49 &0.36 &0.51& SPY: \cite{2007ASPC..372..387N}. {\cite{2017MNRAS.466.1575R}}\\
 WD0453$-$295           &$<-$100&SB2 &0.40 &0.44 &0.36& SPY: {\cite{karl04}}, {\cite{2007ASPC..372..387N}}  \\
 HE0455$-$5315 		&$-$66.18 &SB1 &0.47 & && \\
 WD1013$-$010 		&$-$39.55 &SB1 &0.32 &$\ge 0.62$ &0.44& SPY: {\cite{2005A+A...440.1087N}}, {\cite{2009A+A...505..441K}}  \\
 WD1022+050             &$<-$100 &SB1 &0.37 & $>$0.28 & 1.16 & {\citet{ 1999MNRAS.307..122M}}, {\cite{2005MNRAS.359..648M}}, {\cite{2005A+A...440.1087N}} \\
 HS1102+0934 		&$-$47.41 &SB1 &0.38 &$\ge 0.45$ &0.55 &SPY: {\cite{2005A+A...440.1087N}}, {\cite{2009A+A...505..441K}}. {\cite{2013ApJ...769...66B}}\\
 WD1115+166             &$-$22.28 &SB2 &0.69 & $\ge 0.52$ &30.1& {\cite{2002MNRAS.334..833M}}\\
 WD1124$-$018 		&$<-$100&SB1 &0.49 & && \\
 WD1210+140 		&$<-$100&SB1 &0.33 &$\ge 0.44$ &0.64& SPY: {\cite{2005A+A...440.1087N}}, {\cite{2009A+A...505..441K}} \\
 HS1334+0701 		& $-$6.51 &SB1 & 0.35 & & 0.23:& SPY: {\cite{karl04}}, period ambiguous \\
 WD1349+144 		&$-$53.30 &SB2 &0.53 &0.33 &2.21& SPY: {\cite{2005A+A...440.1087N}}, {\cite{2009A+A...505..441K}} \\
 HE1414$-$0848 		&$<-$100 &SB2 &0.52 &0.74 &0.52& SPY: {\cite{2002A+A...386..957N}}, {\cite{2005MNRAS.359..648M}}, {\cite{2005A+A...440.1087N}} \\
 HE1511$-$0448            &$-$17.85 &SB1 &0.50 &$\ge 0.67$ &3.22& SPY: {\cite{2005A+A...440.1087N}}, {\cite{2009A+A...505..441K}} \\
 WD1824+040 		&$<-$100 &SB1 &0.40 &$\ge 0.73$ &6.27& {\cite{1998ApJ...502..394S}}, {\cite{2005MNRAS.359..648M}}, {\cite{2005A+A...440.1087N}} \\
 WD2020$-$425 		&$<-$100&SB2 &0.81 &0.54 &0.30& SPY: {\cite{2009A+A...505..441K,2007ASPC..372..387N}} \\
 WD2200$-$136 		&$<-$100 &SB2 &0.46 & && \\
 HE2209$-$1444 		&$-$38.47 &SB2 &0.43 &0.72 &0.28& SPY: {\cite{2003A+A...410..663K}}, {\cite{2005MNRAS.359..648M}}, {\cite{2005A+A...440.1087N}} \\
 HS2216+1551 		&$-$99.25 &SB2 &0.64 & && \\
 WD2330$-$212 		&$-$27.79 &SB1 &0.45 & && \\
 WD2336$-$187 		&$-$16.23 &SB2 &0.36 & && \\
 HE2345$-$4810 		&$-$15.62 &SB1 &0.43 & && \\
\hline
\end{tabular}
%\footnotetext[1]{kaka}
\end{center}
\end{table*}

Fig.~\ref{f:HR} 
\begin{figure*}
 \centering
 \includegraphics[width=10cm,angle=-90]{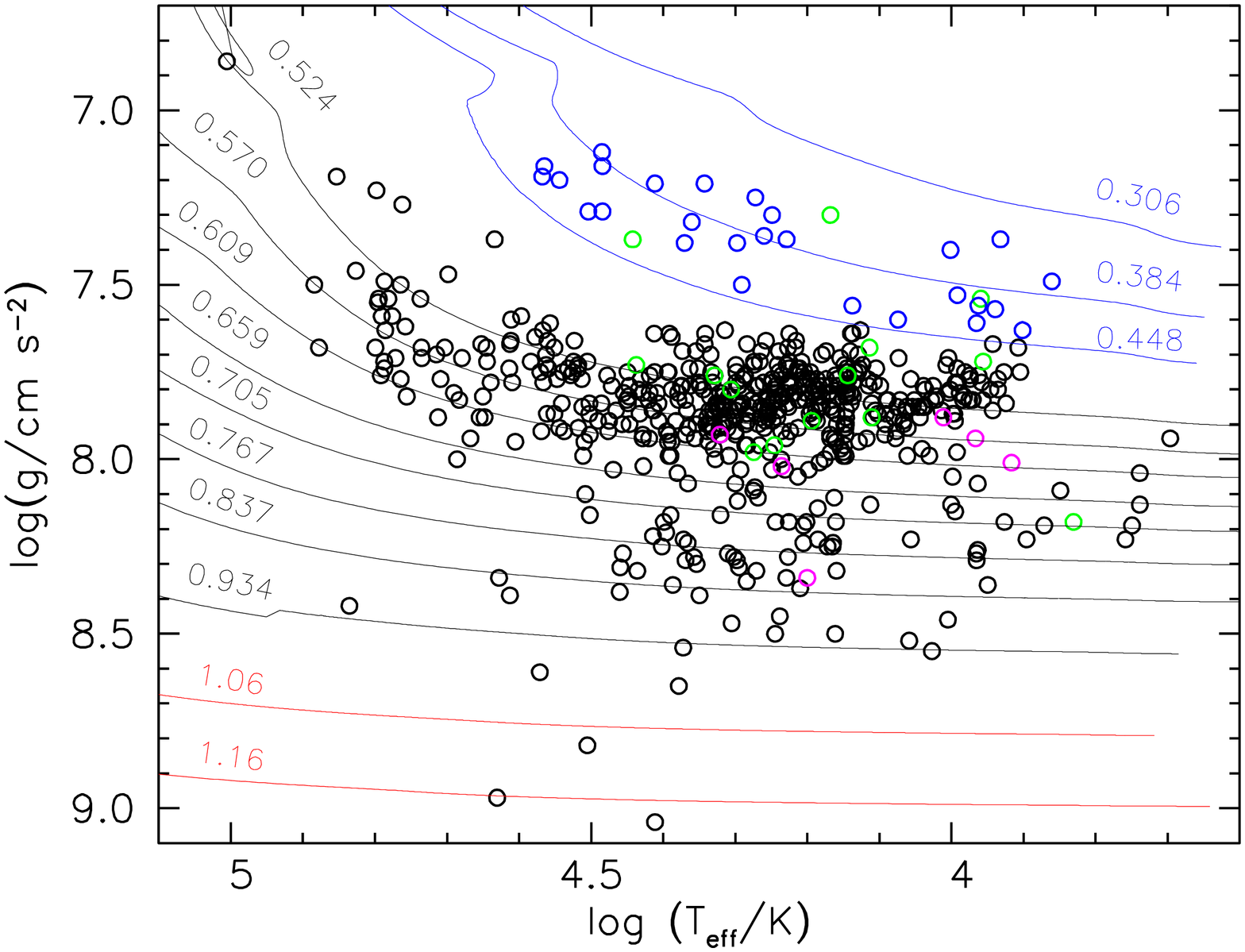}
 \includegraphics[width=10cm,angle=-90]{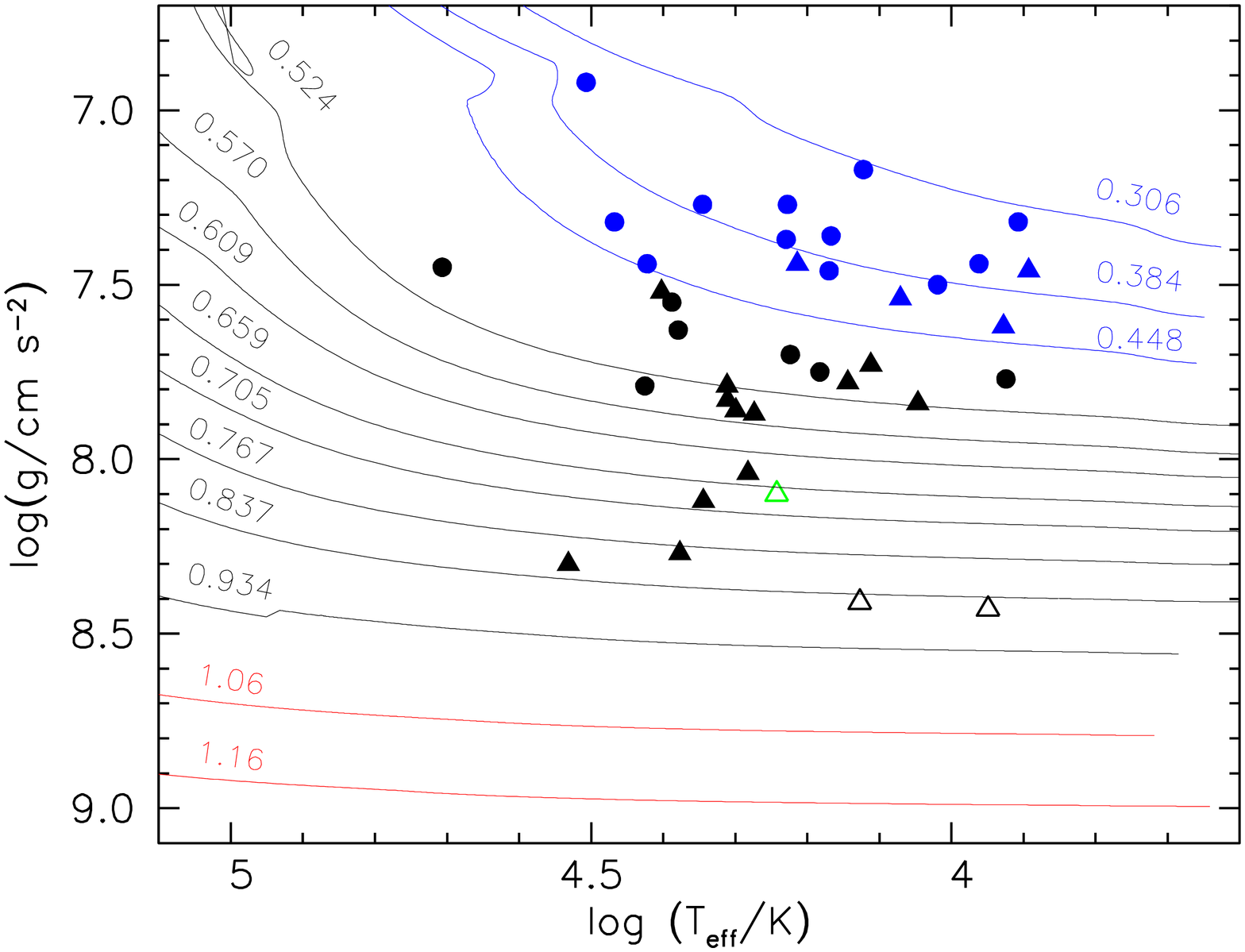}
\caption{Comparison of the SPY sample to evolutionary tracks for He-core  \citep[blue,][]{2007MNRAS.382..779P}, C/O core \citep[black,][]{2010ApJ...717..183R} and O/Ne
core white dwarfs \citep[red,][]{2005A+A...441..689A} in the T$_{\rm eff}$ -- $\log$\,g plane. The tracks are labelled with their mass. Upper panel: Position of all stars in Table~\ref{t:tefflogg} excluding the DD white dwarfs and stars without a mass determination. Magnetic white dwarfs are shown in magenta, those with only one spectrum in Table \ref{t:RV} in green. The hot, low gravity star HE1106$-$0942, the high gravity DA WD0346$-$011, and the magnetic, high gravity DA WD2051$-$208 are off-scale. Lower panel: same for DDs. SB1 systems (19) are shown as filled circles and SB2 systems (19) as triangles; the SB2 WD0135-052 is not shown as we were not able to
determine masses for it.
Filled triangles mark systems that are RV variable, while open ones label SB2 systems that were found by visual inspection of their spectra, but are not considered RV variable (WD0128$-$387 and MCT0136$-$2010) or where only one spectrum was taken (HS0237+1034). The parameters of the brighter component were used. The gravity of HE0225$-$1912 has been offset by -0.05\,dex to improve visibility. \label{f:HR}}
\end{figure*}
shows the distribution of our white dwarf sample in the
\teff/$\log g$ plane with single white dwarfs shown in the upper panel and DDs shown in the lower panel. The
atmospheric values have been taken from \cite{2009A+A...505..441K}. For
\teff$<$12\,500 K the values of \teff/$\log g$ from 1D
models are not accurate and hence we applied the corrections between 1D and 3D models of \cite{2013A+A...559A.104T} for a mixing length parameter ML2/$\alpha$=0.6, which is the value used in the Koester models.

\subsubsection{Double-lined systems}

In Fig.~\ref{f:SB2} we show spectra of 18 of the 20 double-lined DDs found in the survey, excluding only the two DAB white dwarfs, WD0128-387 and WD1115+166, which we discuss below. Double-lined systems are of particular interest since, once fully characterised, they deliver mass ratios and fundamental parameters for both stars and thus are of special importance for the understanding of the prior
evolution of the systems.
We now discuss special cases.

\begin{figure*}
%\fbox{
\includegraphics[width=17cm]{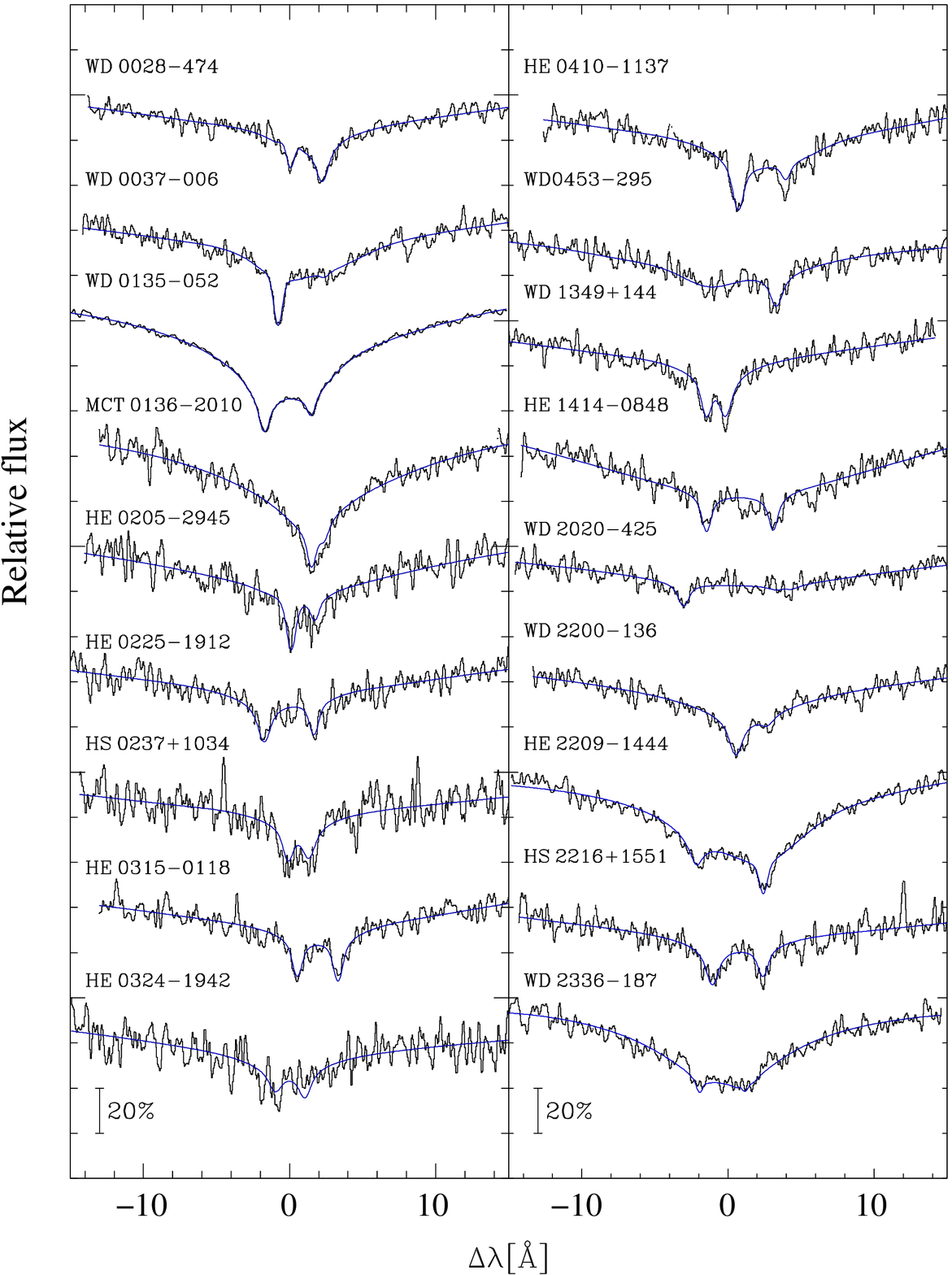}
%}
\caption{\halpha region of 18 of the 20 the double-lined systems detected in the survey. Each system shown displays two \halpha\ cores; the two systems excluded, WD0128-387 and WD1115+166, are DABs and show only a single component at \halpha. The figure shows one observed spectrum for each binary along with the best fitting model spectrum (blue smooth line). The observed spectra were
  smoothed with a Gaussian with 0.15\,\AA\ for display. $\Delta \lambda$ is the
  wavelength offset relative to the laboratory wavelength of \halpha .}
\label{f:SB2}
\end{figure*}

\paragraph{WD\,2020$-$425:} The spectrum shown in Fig.~\ref{f:SB2} appears to 
show only the \halpha core of one object. However, comparison of the two
spectra taken at different epochs (Fig.~\ref{f:wd2020}) demonstrates the
existence of a second component. This white dwarf is in the parameter range
described before (Sect~\ref{s:pec}), in which \halpha has neither an
absorption nor an emission core. WD\,2020$-$425 has parameters, which make it
a possible SN\,Ia progenitor \citep{2005ASPC..334..375N}. Its period is short
enough to let it merge within a Hubble time and the combined mass of both white dwarfs
is very close to the Chandrasekhar limit. However, the lack of a sharp core of
the second, high mass, white dwarf results in relatively high uncertainties, so that a
mass below Chandrasekhar can not be ruled out. 

\begin{figure}
%\fbox{
\includegraphics[height=8.5cm,angle=-90]{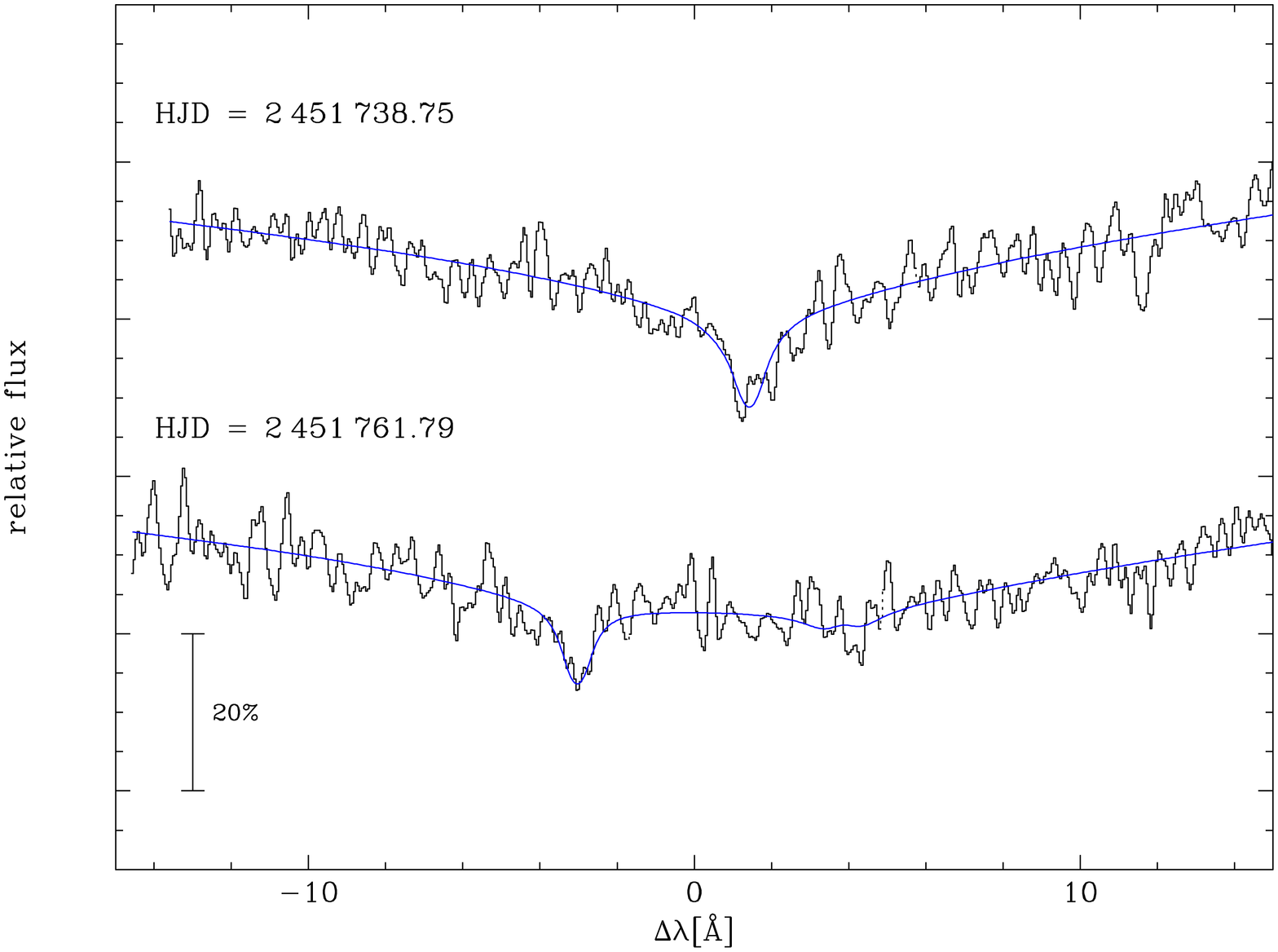}
%}
\caption{\halpha region of the two spectra of WD\,2020$-$425 taken at
  different epochs.}
\label{f:wd2020} 
\end{figure}

\paragraph{HS\,0237+1034} No second epoch survey spectrum exists for 
HS\,0237+1034. However, it is a double-lined DD consisting of two DA white dwarfs 
(Fig.~\ref{f:SB2}).

\paragraph{MCT\,0136$-$2010:} Two spectra were taken for this target. The RV 
shift measured between the two spectra is not significant. The reason for us
to classify it as DD is its double-lined spectrum. This is not too obvious at
a first glance at Fig~\ref{f:SB2}. However, the two spectra taken at different
epochs (Fig.~\ref{f:mct0136}) are clearly different and the $\chi^2$
difference between a fit with and without a second component is 130.5, highly
significant.  Figure~\ref{f:mct0136} shows the \halpha profile fitted with a
single model.

The small RV offsets in both epochs makes a small RV amplitude and thus a long
period likely. Although it is thus unlikely to qualify as SN\,Ia progenitor,
it appears to have a high mass. The fundamental parameters derived by
\citet{2009A+A...505..441K} from a single-lined fit are $\teff=8416\,K$ and
$\log g = 8.405$, making it an interesting object for further follow-up.

\begin{figure}
%\fbox{
\includegraphics[height=8.5cm,angle=-90]{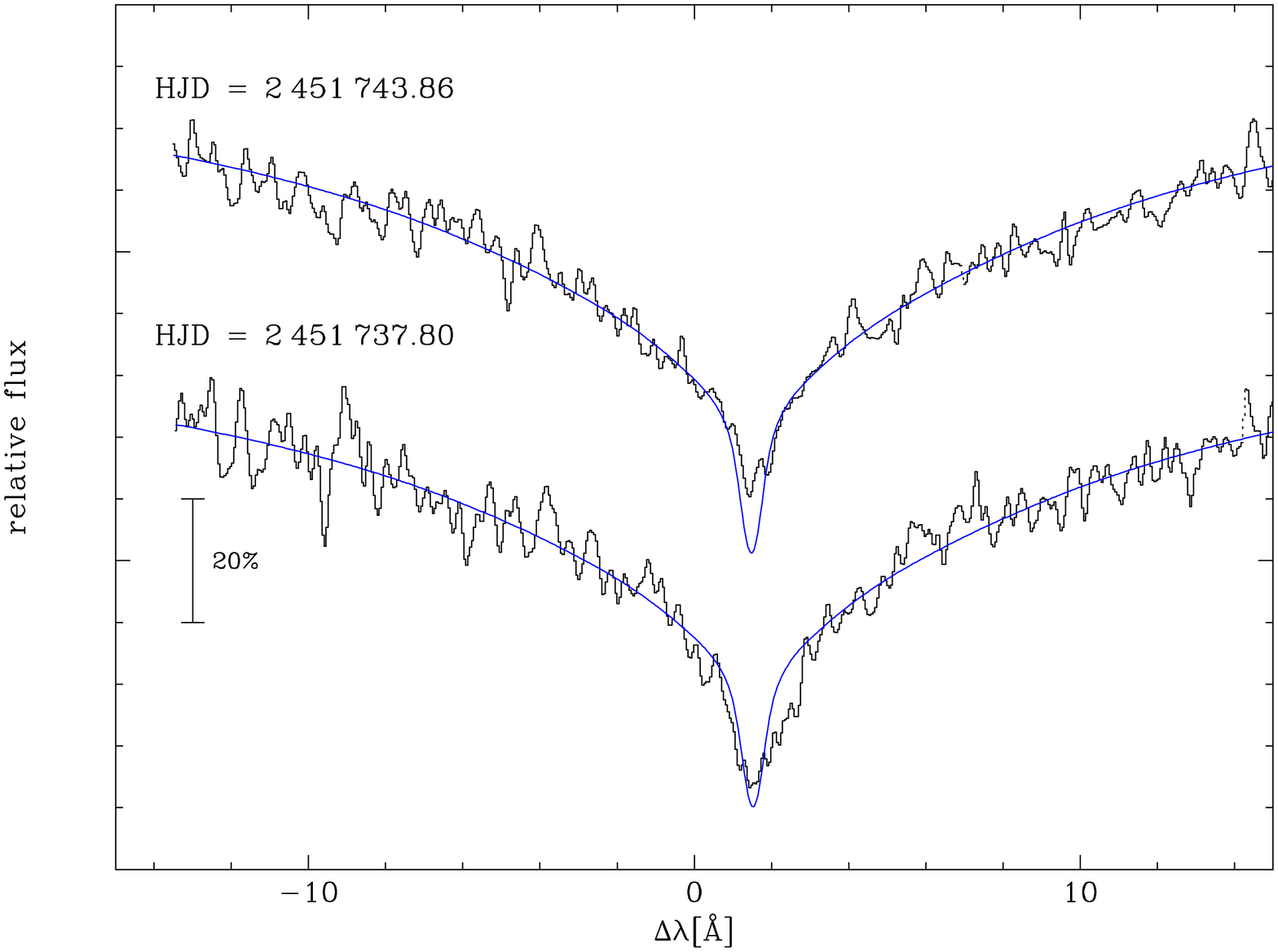}
%}
\caption{\halpha region of the two spectra of MCT\,0136$-$2010 taken at
  different epochs fitted assuming a single-lined spectrum.}
\label{f:mct0136} 
\end{figure}

\paragraph{Spectral types DAB and DAO:}

Two of the DABs (WD\,0453$-$295 and WD\,1115+166) show large RV variations and are clearly short period DDs. The spectrum of WD\,0453$-$295 shows two \halpha\ components making this a DA+DBA binary. No second \halpha\ component is visible in WD\,1115+166 and it is thus a DA+DB binary, in agreement with the detection of the binary period and anti-phased motion in the hydrogen and helium lines by \citet{2002MNRAS.334..833M}. The RV shift measured for WD\,0128$-$387 has barely $1\,\sigma$ significance. However, we noted that in our
spectra the He~I lines of the DB component are blueshifted by $60\ldots 70\kms$ relative to the Balmer lines. The Balmer lines show a weaker blueshifted component, too. This could be explained by gravitational redshift,
if the DA were the more massive white dwarf in the system. However, according to the
model atmosphere analysis of \citet{1994ApJ...429..369W} the DB has higher
gravity ($\log g = 9.0$) than the DA ($\log g = 8.5$) thus ruling out
gravitational redshift as explanation. We conclude that, although the evidence is
somewhat circumstantial, WD\,0128$-$387 is a close DD system consisting of a
DA and a DBA. No significant RV variations were detected for the two DAOs HE\,1106$-$0942 and WD\,1305$-$017
included in the SPY sample.

\subsection{Notes on individual objects}

%\paragraph{WD\,0135$-$052:}
%This is a well known double-lined DD %\citep{1988ApJ...334..947S}, which
%somehow managed to sneak into the SPY input
%catalogue. \citet{1988ApJ...334..947S} determined an 5orbital period of 1.56
%days.

\paragraph{WD\,1845$+$019:} 
This object is listed as a white dwarf with a cool companion candidate in
\cite{2007AJ....134...26H}. The authors find this white dwarf has a near-IR
excess in Two Micron All Sky Survey
\citep[2MASS,][]{2006AJ....131.1163S}. \cite{1999MNRAS.307..122M} found
spectroscopic evidence for a long-period, red companion inferred from the
presence of narrow, stationary emission near the core of H$\alpha$.  The SPY
spectra show similar emission at H$\alpha$, whereas \cite{1995ApJ...448..305S}
did not report any sign of such emission from spectra which also
covered H$\alpha$.  It is not known whether the close neighbour star described
by \cite{2005AJ....130.1221D} is gravitationally bound to the white dwarf and whether
it is the same as the companion star inferred by \cite{1999MNRAS.307..122M}.
This object was erroneously listed in \cite{2009A+A...505..441K} as a double
degenerate. We measured RVs modelling \halpha as a composite of a DA spectrum and a
Gaussian emission. Our measurements have small error bars and are consistent
with constant velocities.
%This object has been then removed from the main table, since only
%isolated white dwarfs and DDs are listed there.

\paragraph{WD\,2359$-$434:} The line cores of this white dwarf have
a strange flat profile with a weak narrow component already reported in
\citet{1998A+A...338..612K}. \cite{1998A+A...338..612K} speculated that
WD\,2359$-$434 is magnetic and that what they see is only the unshifted
component of the Zeeman triplet, with the other components shifted outside
their observed spectral range or smeared out due to the inhomogeneity of the
field. \citet{2004A+A...423.1081A} included this object in their
spectropolarometric observations and discovered a weak magnetic field of
3\,kG. However, our spectra do not show any shifted Zeeman components. A
complicated field structure could smear out the shifted components, but a
3\,kG field would be too weak to produce smearing over the necessary
wavelength range.

\cite{2012A+A...545A..30L} reanalysed UVES and FORS data of this object and
detected a non-uniform magnetic field in the range 3.1-4.1kG. On the other
hand, \cite{2012MNRAS.425.1394K} obtained a magnetic field of 9.8kG from
spectropolarometric observations.

Infrared spectroscopy puts very tight constraints on any cool main sequence or
substellar companion \citep{2005MNRAS.357.1049D}. Even brown dwarfs with a
spectral type earlier than T8 (corresponding to a mass of $0.05\,\msun$ for an
age of 10\,Gyr) can be ruled out. No sign of an infrared excess produced by
warm circumstellar dust can be seen in the J,H and K bands.

For our RV measurements we modelled the flat bottom line profiles assuming fast
rotation and broad Lorentzians, i.e.\ the narrow component was ignored. Our
assumption of fast rotation is only an attempt to model the profile and does
not imply that WD\,2359$-$434 is really a fast rotator. Our spectra, taken
four days apart, do not show a significant RV shift.

\subsubsection{Magnetic white dwarfs}

Ten magnetic white dwarfs have been identified in the SPY survey by Zeeman
splitting. WD\,0058$-$044 and WD\,0239$+$109 were already published by
\cite{2001A+A...378..556K}. HS\,0051$+$1145, HE\,1233$-$0519 and
WD\,2051$-$208 were first identified in SPY and details can be found in
\cite{2009A+A...505..441K}.  Three other magnetic white dwarfs
(WD\,0257+080, WD\,1953$-$011 and WD\,2105$-$820) were studied by SPY,
confirming their magnetic nature, but were already published in previous
works. More information on these objects can be found in
\cite{2009A+A...505..441K}. Further to these objects in this work we have also
identified two other magnetic white dwarfs that were already published
(WD\,1300$-$098 and WD\,1350$-$090). More details on these two objects are
given below.

\paragraph{WD\,1300$-$098:} 
The SPY spectra of this object showed Zeeman splitting in the Balmer lines
\citep{2009A+A...505..441K}. This was later confirmed in
\cite{2011ApJ...743..138G}.

\paragraph{WD\,1350$-$090:} 
This is a well known magnetic white dwarf which was found through
spectropolarimetry to have a weak magnetic field of 85kG
\citep{1994ApJ...423L..63S}. \cite{2005PASP..117.1378W} studied the NIR values
of this object looking for companions, but did not find one.  The Zeeman
splitting in the Balmer lines of the SPY spectrum of this white dwarf was first
reported in \citet{2001A+A...378..556K}.

\subsection{Comparison with \citet{2017MNRAS.467.1414M}}

\citet{2017MNRAS.467.1414M} presented a similar analysis of a subsample of 439 DAs out of the 615 SPY DAs analysed by \citet{2009A+A...505..441K} %Koester et al. 2009
%\SMo{\it 615-135 = 480?)} 
We start from 643 stars of which 18 have one spectrum only, leaving us with 625 stars. Hence, the sample of \citet{2017MNRAS.467.1414M} is a subset of ours.  

The criteria employed to identify radial velocity variables differ significantly between our two studies. While we use the $\log p < -4$ detection threshold (see Sect. \ref{s:results}), \citet{2017MNRAS.467.1414M} prefer $\Delta RV_{\rm max} > 15$ km\,s$^{-1}$ to identify a radial velocity variable star and  $10 < \Delta RV_{\rm max} < 15$ km\,s$^{-1}$ for candidates.
We confirm all of the \citet{2017MNRAS.467.1414M}'s detections except for HE0516$-$1804 ($\log p = -0.72$), HS2046+0044 ($\log p = -2.54$), and WD0032$-$317, although as discussed earlier, this latter object could well prove to be a DD 
given its false alarm probability of $\log p = -3.74$.

We find ten stars to be radial velocity variable which are not listed in \citet{2017MNRAS.467.1414M}: the known SB2 systems HE0315$-$0118, WD0453-295 and WD1115+0934, the known SB1 systems WD0216+143, WD0101+048, WD1022+050, and HE1511-0448 as well as the SB1s HE0455-5315, HS1334+0701, and HE2345-4810. These objects are probably among their rejects. In addition, visual inspection of the UVES spectra allowed us to identify %WD0135-052, 
MCT0136-2010 and HS0237+1034 as SB2 systems (see Fig. \ref{f:SB2}). %Figure 7

Amongst \citet{2017MNRAS.467.1414M}'s candidate RV variable objects ($10 < \Delta RV_{\rm max} < 15$km\,s$^{-1}$), we confirm that HS2216+1551 is a radial velocity variable SB2 system, however, all 15 remaining 
candidates listed in their Table~1 are rejected using our criterion.

In summary, we reject 17 out of the 43 DDs of \citet{2017MNRAS.467.1414M},
confirming only 25 of them. Of these, 7 have system parameters published by the SPY consortium before, and 3 from other sources. From our full sample we identified 39 DDs including 20 SB2 systems.
We identify 12 SB1 and 13 SB2 systems in the \citet{2017MNRAS.467.1414M} subsample. However, the overall DD fraction in both samples is the same at 6\%, if the same variability test is applied. 

The $\Delta RV_{max}$ statistic employed by \citet{2017MNRAS.467.1414M}
has two intrinsic weaknesses which could perhaps have contributed to the differences between our studies. First it takes no account of the number of spectra -- the expected distribution of $\Delta RV_{max}$ from purely statistical fluctuations always shifts towards larger values as the number of spectra considered increases. For instance, we find that a threshold that is only exceeded by chance $1$\% of the time with two spectra, is breached almost three times as often with three spectra, a purely statistical and spurious effect. There are 
perhaps indications in \citet{2017MNRAS.467.1414M}'s results of the influence this effect: three of their 16 candidates have three or more spectra in SPY, whereas only one such case would have been expected by random chance. The second problem with $\Delta RV_{max}$ is that it  takes no account of varying statistical uncertainties between spectra.
These issues are why we employed \citet{2000MNRAS.319..305M}'s
$\chi^2$-based criterion.

%WD1115+0934  log p=-22.28   SB2, this is missing in your Fig. 2, there are only 19 Halpha profiles shown, not all 20 as claimed

%WD0101+048 log p=-17.5!
%WD0135-052 log p=-0.48, but SB2! Fig. 2 , not easy to see
%MCT0136-2010 log p=-039, but SB2! s. Fig. 2 clear cut case
%WD0216+143   log p=-6.75
%HS0237+1034  log p=0.0 but SB2! Fig.2
%WD0453-295   log p<-100! SB2! Fig. 2, not easy to see, but is published in Napiwotzki et al. 2007
%HE0455-5315  log p=-66.18!!
%WD1022+050   log p<-100!  this is in Morales-Rueda et al. 2005 and Nelemans et al. 2005
%WD1115+0934  log p=-22.28   SB2, this is missing in your Fig. 2, there are only 19 Halpha profiles shown, not all 20 as claimed
%HE1511-0448  log p=-17.85 also Nelemans et al. 2005
%HE2345-4810  log p=-15.62
%Hence, they did not spot 5! SB2 systems (1 published). 
%Hence, HE2148-3857 is not in the sample although listed in Table B.1. for the single spectrum.
%The reason why  WD0101+048 and WD0135-052 are not in MH may be that they are among the 135 rejects.

\section{Discussion}
\label{s:discussion}

We have presented RV measurements of a set of 643 DA white dwarfs observed for the SPY programme and remaining after 46 binaries with cool companions were removed from the original sample of 689. We obtained two or more spectra for 625 of these targets, 632 including targets covered by \citet{2000MNRAS.319..305M}. This sample contains 39 DDs (20 of the SB2 type, 19 of the SB1 type), 35 of them new discoveries.  The directly detected rate of DDs in SPY is thus $6.2\pm1.0$\% (uncertainty is $1\sigma$ from the binomial distribution).
There are likely to be an additional 5 or 6 DDs in our sample that we did not detect, given typical periods of a day and our mean detection probability (Fig.~\ref{f:detection}), but a deduction of the space density of DDs requires accounting for much more than just period selection, as we discuss below in section~\ref{s:selection}, so we prefer for simplicity to quote just the 
numbers as observed. 

\subsection{Dependence of DD numbers on mass}
The SPY sample is large enough to see
some significant variations with mass.
The limiting mass for a He white dwarf is set by the minimum mass for a helium core
flash which increases from $0.43$ to $0.49\,\msun$ \citep{1993ApJ...419..596D,1994ApJ...426..612S} with decreasing metallicity, although it is common to adopt a value of
$0.45\,\msun$.  White dwarfs with higher masses usually have cores composed of
carbon and oxygen. There are scenarios that lead to the formation of low-mass
``hybrid'' white dwarfs with C/O cores and thick helium mantles \cite[$\sim 0.1\,\msun$][]{1985ApJS...58..661I}. Some white dwarfs with masses around $0.45\,\msun$
may also have evolved via the extreme horizontal branch (EHB), but these
objects are also considered to be the product of close binary evolution
\citep[e.g.\ ][]{2009ARA+A..47..211H,2015A&A...576A..44K,2016PASP..128h2001H}.  So low-mass white dwarfs may not
always be helium core white dwarfs, although in all cases evolution within a
binary is needed to model the early mass loss that prevents evolution towards
a normal C/O WD.

Fig.~\ref{f:mass_distrib}
\begin{figure}
\includegraphics[width=8.5cm]{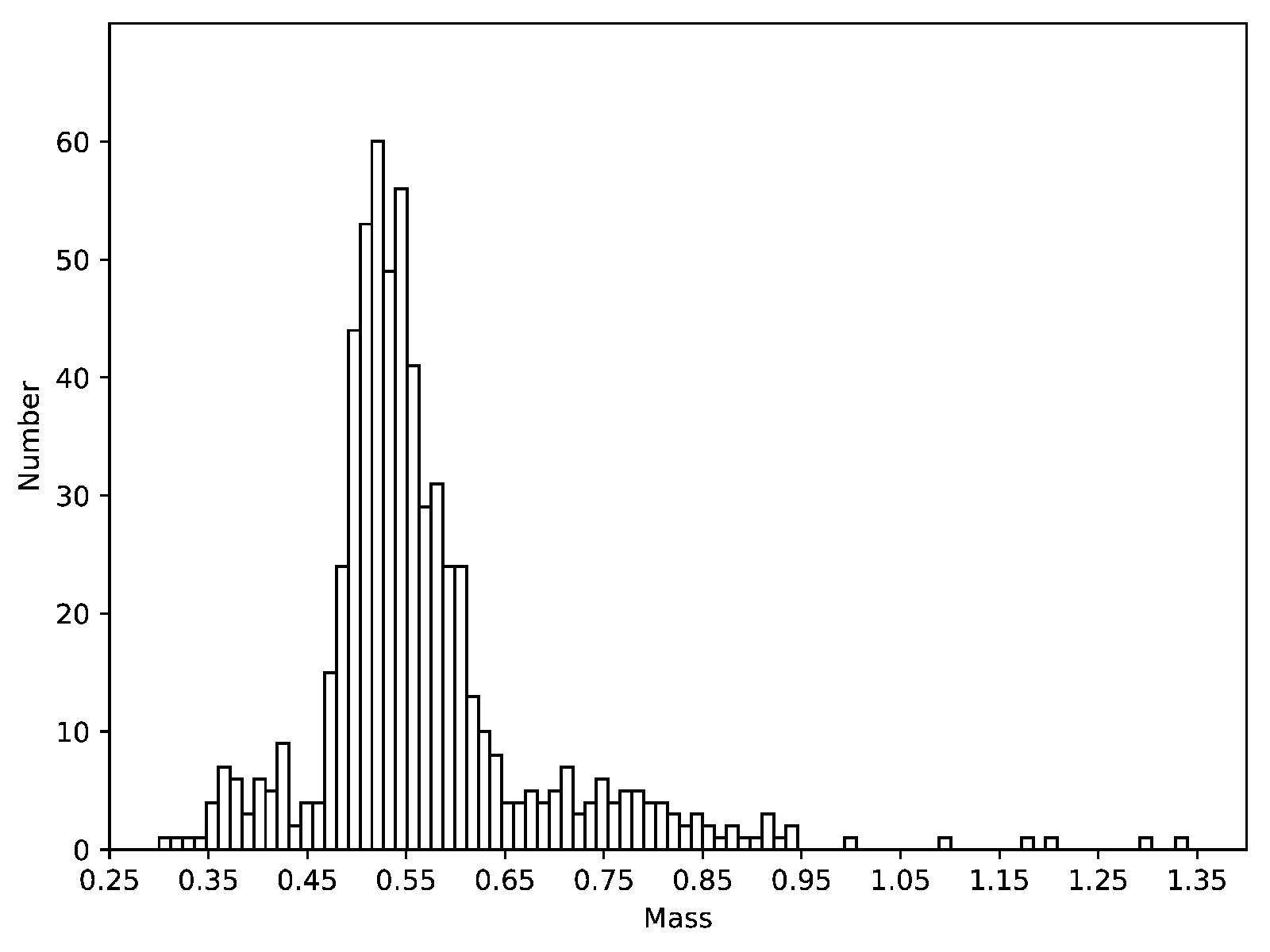}
\caption{Mass distribution of the SPY sample.} 
\label{f:mass_distrib}
\end{figure}
displays the mass distribution of 615 DA white dwarfs of the SPY sample for which we could derive masses. The distribution may be four populations as suggested by \citet[][cf. their Fig. 8]{2007MNRAS.375.1315K}. To examine how DD numbers vary with mass, we consider those targets that (a) have masses in Table~\ref{t:tefflogg} and (b) have two of more spectra and are thus sensitive to DDs. We arrive at the numbers listed in Table~\ref{t:mass_groups}.}
\begin{table}
\caption{Breakdown of targets with two or more spectra into four mutually exclusive sets defined by mass. (Exception:
HS0237+1034 has only one spectrum, but is double-lined.)
Notation: "[" or "]" -- interval includes the limiting value; "(" or ")" -- interval excludes the limiting value.\label{t:mass_groups}}
\begin{tabular}{l|r|r|r|r}
\hline
Type & I  & II  &  III & IV\\
  & $[0,0.45]$  & $(0.45, 0.65]$  & $(0.65,1.00]$ & $(1.00,1.40]$ \\ 
\hline
all & 44 & 478  & 83 & 6 \\   
SB1 & 12 &  7  & 0 & 0  \\
SB2 & 4 & 9  & 6 & 0 \\
\hline
\end{tabular}

\end{table}
All the DDs bar WD0135-052, for which we could not determine a mass from SPY data,
contribute to this table, i.e. 38 of the 39 DDs in SPY.
We identify the 44 targets with primary masses $M_1 \le 0.45\,\msun$ (category I) as He white dwarfs. The dominant population II ($0.45 < M_1 \le 0.65\,\msun$) is identified with C/O-core white dwarfs. A third population of higher masses  ($0.65 < M_1 \le 1.0\,\msun$) should also consist of C/O-core white dwarfs. A few white dwarfs have masses in excess of $1.0\,\msun$ (population IV) and could include O/Ne-core white dwarfs and mergers in addition to C/O-core stars.

Of the 44 white dwarfs that have a mass below
$0.45\,\msun$, 16 show radial velocity variations indicating that
they are double degenerates. In terms of detection probability, a set of $0.4$+$0.4\,\msun$ DDs is comparable to (albeit a little below) the line 
corresponding to a companion mass of $0.55\,\msun$ in Fig.~\ref{f:detection}.
Thus, assuming periods of order a day, if all 44 were DDs, we would have expected to detect $\sim 37$ of them as such, $\sim 33$ if all companions were ELMs or low-mass main-sequence stars, and still $\sim 20$ to $26$ for $0.05\,\msun$ brown-dwarf companions with periods comparable to those discovered to date \citep{2017MNRAS.471..976P}. The observed number of 16 suggests that some of remaining 28 non-detections could be single, or be harbouring very low mass companions. 
It has been suggested that single low mass white dwarfs result from envelope loss induced by capture and spiral-in of
planetary companions \citep{1998A+A...335L..85N}, or are the cores of red giants stripped by explosions of their  companions as supernovae 
\citep{2009A+A...493.1081J}, or are the products of
mergers of components of cataclysmic binaries induced by frictional
angular momentum loss following nova eruptions \citep{2017MNRAS.466L..63Z}.
All possibilities are interesting and these systems are worth further investigation.

The detection of 16 DDs amongst the 44 He white dwarf primaries is a fraction of $36.4\pm7.3$\%. By comparison, the 22 DDs in the remaining 567 systems with $M_1 > 0.45\,\msun$, a fraction of 
just $3.9\pm0.8$\%, is a steep drop. For the dominant population of C/O white dwarfs (set II) the binary fraction is even lower (3.3\%), while among the C/O white dwarfs of set III 7.2\% are binaries. No binary was found among the white dwarfs with masses exceeding  1\,M$_\odot$ (set 4). The He white dwarf DDs of set I divide 12:4 in favour of single-lined systems, while those of higher mass are skewed 7:15 towards double-lined systems (sets II-IV), in particular for set III, in which all six binary systems are double-lined. The dominant population of C/O white dwarfs (set II) divide almost equally. This is perhaps because low mass white dwarfs are large and will tend to dominate over their companions, but it could also be a consequence of the formation paths that result in these systems. There is useful potential here to constrain evolutionary models.

\subsection{Selection effects in the SPY sample.}

\label{s:selection}

The current sample of DDs has emerged from searches that often started with mass-dependent selection, following successful searches of the 1990s which concentrated upon low mass white dwarfs \citep[e.g.][]{1995MNRAS.275..828M}. The most extreme example of this strategy is the ELM survey \citep{2016ApJ...818..155B} which uncovered targets so low in mass that they were not even recognised as white dwarfs when the SPY survey was undertaken. Such surveys are biassed precisely against the high mass systems of most interest as Type~Ia progenitors, except to the extent that some may harbour high mass white dwarfs hidden in the glare of dominant low mass primary stars. They also focus on the extreme low mass wing of the white dwarf mass distribution, which is untypical of most systems. SPY is by far the most sensitive survey without mass selection built in from the start. Precursors to SPY were \cite{1990ApJ...365L..13B} and \cite{2000MNRAS.319..305M} who surveyed 54 and 71 white dwarfs on 2--4\,m-class telescopes, uncovering a handful of candidates, although the second of these studies emerged from searches that concentrated in the main upon low mass targets. The Sloan Digital Sky Survey (SDSS), although not designed for radial velocity work, serendipitously provides a very different type of constraint, providing a much larger sample, but also much looser radial velocity constraints. The SDSS is perhaps somewhat superior to SPY for finding massive, short-period Type~Ia progenitors \citep{2017MNRAS.468.2910B}, but SPY is far more sensitive to the bulk population of longer period and lower mass DDs.  Moreover, SPY is particularly efficient in finding double-lined systems because of its high spectral resolution. At lower resolution such as the SDSS spectra, it is far more difficult because the components' line profiles are not resolved and the orbital motion will broaden the combined line profile rather than shifting the unresolved line centre. This is probably why double-lined DDs have not been found directly by the SDSS project. In addition visual inspection of SPY spectra allowed SB2 systems to be discovered that were considered RV non-variable by the detection criterion and, therefore, probably have long orbital periods. Hence, SPY  will enable a more complete and reliable calibration of theoretical models. The two surveys are in many ways complementary to each other \citep{2018MNRAS.476.2584M}, as can be judged from the detection probability plot for SDSS \citep[Fig.~4; ][]{2017MNRAS.468.2910B}, comparable to Fig.~\ref{f:detection} in this paper. The SDSS detection rate falls below $50$\% for $P > 0.2\,$days, whereas SPY extends over 100 times longer, only dropping below $50$\% for $P > 30\,$days. The numbers of DD detections in each survey corroborate this: the 6396 SDSS targets studied by \cite{2017MNRAS.468.2910B} yielded 15 DDs, while we have found 39 DDs in 625 SPY targets. One can deduce $\sim 400$ other DDs in the SDSS sample that have yet to be detected as such.  

The SPY sample was selected without regard to mass, but it inherits the selection effects of the input catalogues it was derived from, and accounting for these will be crucial 
when making comparisons to theoretical population models.
A non-exhaustive list of such effects is: (a) period-dependent selection (Fig.~\ref{f:detection}), (b) selection by galactic location (Fig.~\ref{f:ait}), (c) colour selection, (d) selection by kinematics via
proper motion surveys, and (e) selection by magnitude limits. All of these can be translated to a greater or lesser extent into mass-dependent selection functions. For instance, selection by temperature favours high mass targets since it takes longer for high mass white dwarfs to cool to $10\,000\,$K than it does low mass white dwarfs, owing to a combination of available thermal energy and surface area. On the contrary, magnitude-limited surveys strongly favour low mass white dwarfs because of their greater size. There are also subtler effects such as mass-dependent kinematics \citep{2012MNRAS.426..427W} which can lead to both proper motions and scale-heights varying with white dwarf mass. The correct way to allow for these effects is to impose them as closely as possible upon theoretically-derived samples rather than attempt to remove them from the observations. To this end we include a compilation of temperatures and gravities of the SPY sample along with Gaia IDs in Appendix~\ref{s:fundamental} to facilitate future calculations. SPY-like samples can be generated from theoretical populations by selecting targets of similar colours and magnitudes and imposing in addition a similar selection by Galactic latitude. A full calculation would also simulate the detection of DDs along the lines of \cite{2019MNRAS.482.3656R}. Such model-dependent calculations are far beyond our scope here where we aim instead to try to convey a feel for the level of the selection effects affecting the SPY sample, beyond the obvious ones of Galactic latitude and period discussed earlier.

\begin{figure*}
\hspace*{\fill}
\includegraphics[width=9cm]{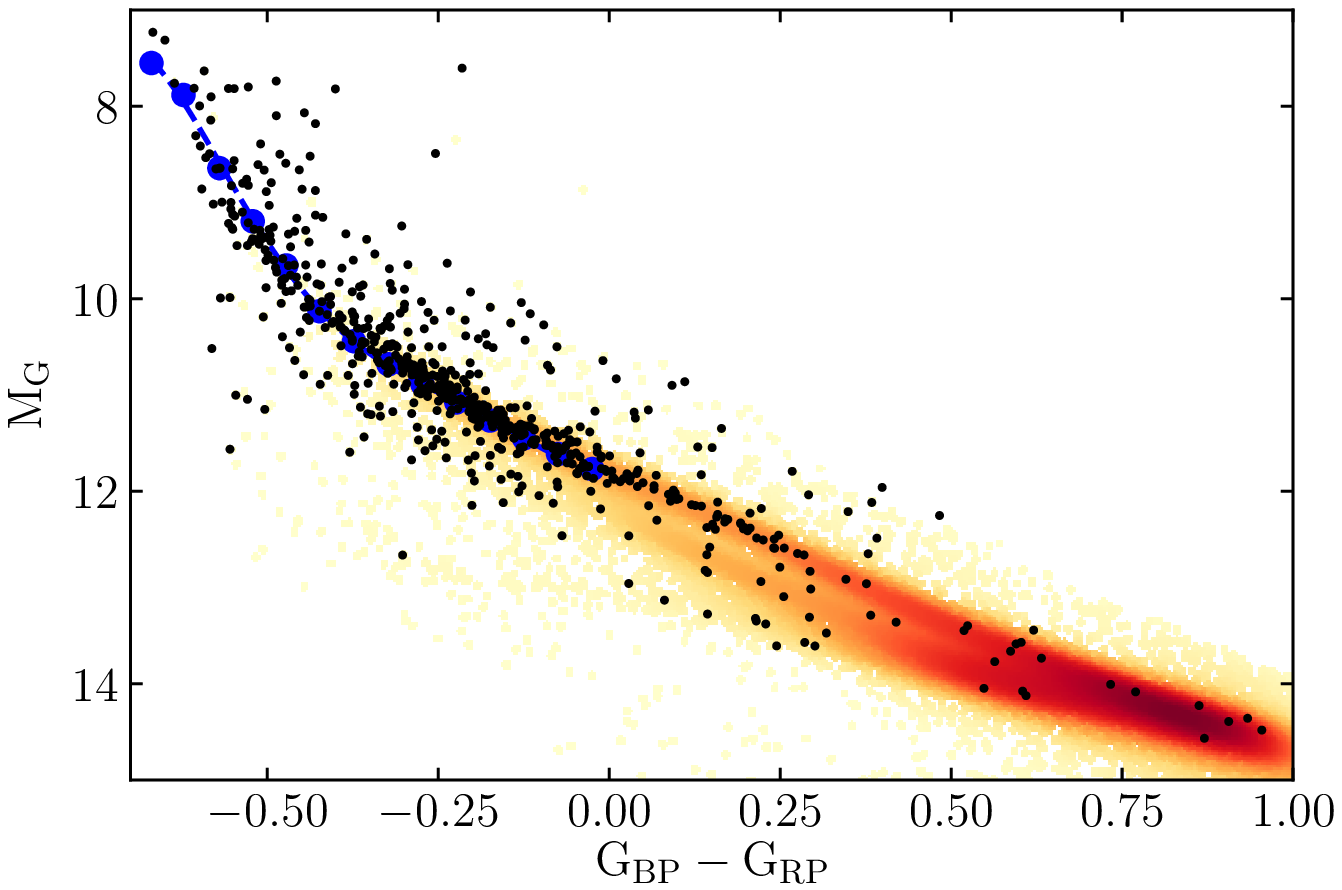}
\hspace*{\fill}
\includegraphics[width=9cm]{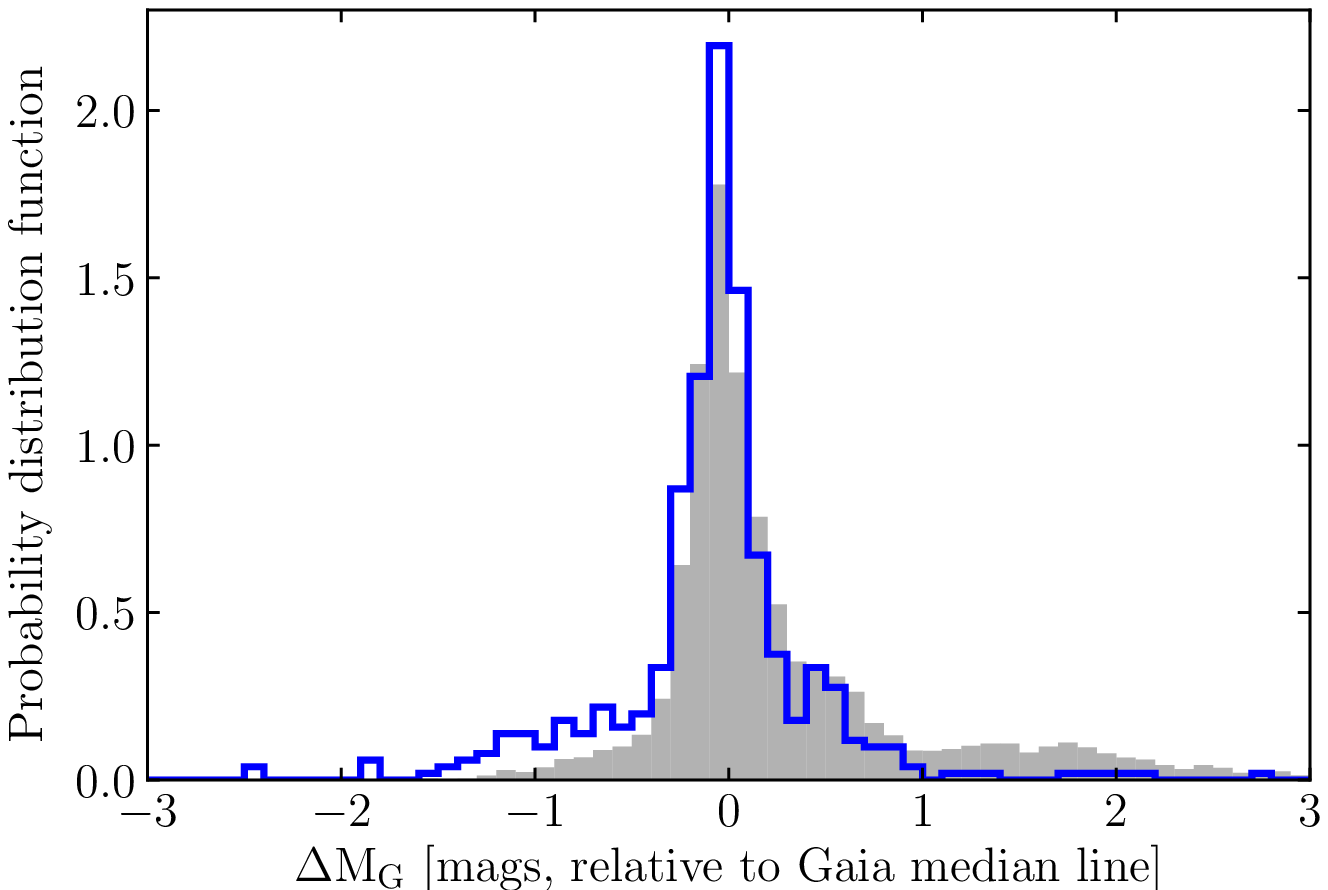}
\hspace*{\fill}
\caption{\emph{Left:} The SPY sample (black dots) plotted in a Hertzsprung-Russell diagram (Gaia absolute magnitude versus Gaia colour) on top of a background of $11\,561$ white dwarf candidates selected from the Gaia-derived sample of \protect\cite{2019MNRAS.482.4570G} and colour-coded by plot density.  The parallax $\pi$ and parallax error $\sigma_\pi$ were restricted by $\pi > 10\,$mas and $\pi > 5\sigma_\pi$ to give a near volume-limited sample to $d = 100\,$pc, although there is still incompleteness in the lower part of the plot since $\mathrm{M}_\mathrm{G} > 14$ at $100\,$pc corresponds to $G > 19$. The large blue dots on the left mark the main cooling track of the Gaia white dwarfs, avoiding the bifurcation that sets in at $\mathrm{G}_\mathrm{BP}-\mathrm{G}_\mathrm{RP} > 0$. The dashed line represents a polynomial fitted to these that was then subtracted from the Gaia and SPY targets to derive differential absolute magnitude histograms. \emph{Right:} Histograms of the differential magnitudes for 
Gaia (solid grey) and SPY (outline blue). The SPY sample has relatively more intrinsically bright objects than the Gaia sample, and thus fewer massive white dwarfs. \label{f:bias}}
\end{figure*}

Given the heterogeneous origins of the SPY input catalogue from a multitude of proper-motion, colour and objective prism surveys as well as serendipitous discoveries, it is close to impossible to unravel the
selection effects of the original survey papers. Instead, we take advantage of the Gaia DR2 survey \citep{2018A&A...616A...1G} to quantify the extent of mass-dependent bias within the SPY sample selection as shown in Fig.~\ref{f:bias}.
We do so by choosing an approximately volume-limited sub-sample from Gaia as an approximation of what is naturally produced by population synthesis calculations. Unsurprisingly,
the SPY sample\footnote{Apart from the following seven SPY targets that have either no
parallax or no colour in Gaia: PG0922+162A, PG0922+162B, WD1015+076,
HE1117$-$0222, WD1121+216, WD1147+255 and WD1214+032}
is very differently distributed to the Gaia sample as shown in the Hertzsprung-Russell diagram on the left of Fig.~\ref{f:bias}. Above all, it is strongly skewed towards blue colours and high temperatures, reflecting the origins of the sample. The temperature selection automatically implies mass-dependent selection because of differential white dwarf cooling rates, as discussed above.

To judge the significance of additional mass-dependent selection biases,
particularly that associated with the magnitude-limited input catalogues, we next look at the distribution of magnitudes \emph{across} the main white dwarf cooling track, restricting ourselves to Gaia colours
$\mathrm{G}_\mathrm{BP}-\mathrm{G}_\mathrm{RP} < 0$ to avoid a region of bifurcation in the Gaia H-R diagram believed to be associated with helium-dominated atmospheres \citep{2019MNRAS.482.4570G}. (The differential effect of atmospheric composition in this region can be seen directly in Fig.~\ref{f:bias} where our hydrogen-dominated DA sample clusters mainly along the upper of the two branches once $\mathrm{G}_\mathrm{BP}-\mathrm{G}_\mathrm{RP} > 0$.) The restriction in colour builds differential cooling mass-selection into the Gaia-selected reference sample too, so we are looking here for additional effects such as kinematic and magnitude bias.
The key result is in the right-hand panel of Fig.~\ref{f:bias} which shows that over-luminous targets ($\Delta \mathrm{M}_\mathrm{G} < -0.5$) are over-represented in SPY compared to a near-volume-limited Gaia sample, whereas faint systems ($\Delta \mathrm{M}_\mathrm{G} > +1$) are under-represented. This is as one would expect of a magnitude-limited sample, but the bias is slight compared to samples chosen for low mass, which are equivalent to applying the restriction $\Delta \mathrm{M}_\mathrm{G} < -0.5$ from the start. Translating Fig.~\ref{f:bias} into a mass-bias is complex, since, while we can assume that $\Delta \mathrm{M}_\mathrm{G} > 0$ targets are more massive than average, double-lined systems with equal contributions from both components get a $-0.75$ magnitude boost, and could be  reasonably massive but still end with $\Delta \mathrm{M}_\mathrm{G} < 0$. Such calculations are
beyond our scope here as they would need to be made separately for any model population. The key point perhaps is that although the biases in the SPY sample are small compared to others in the literature, they are still significant and must be accounted for in any calculation of intrinsic DD fractions and merger rates.

\begin{figure}
\includegraphics[width=8.5cm]{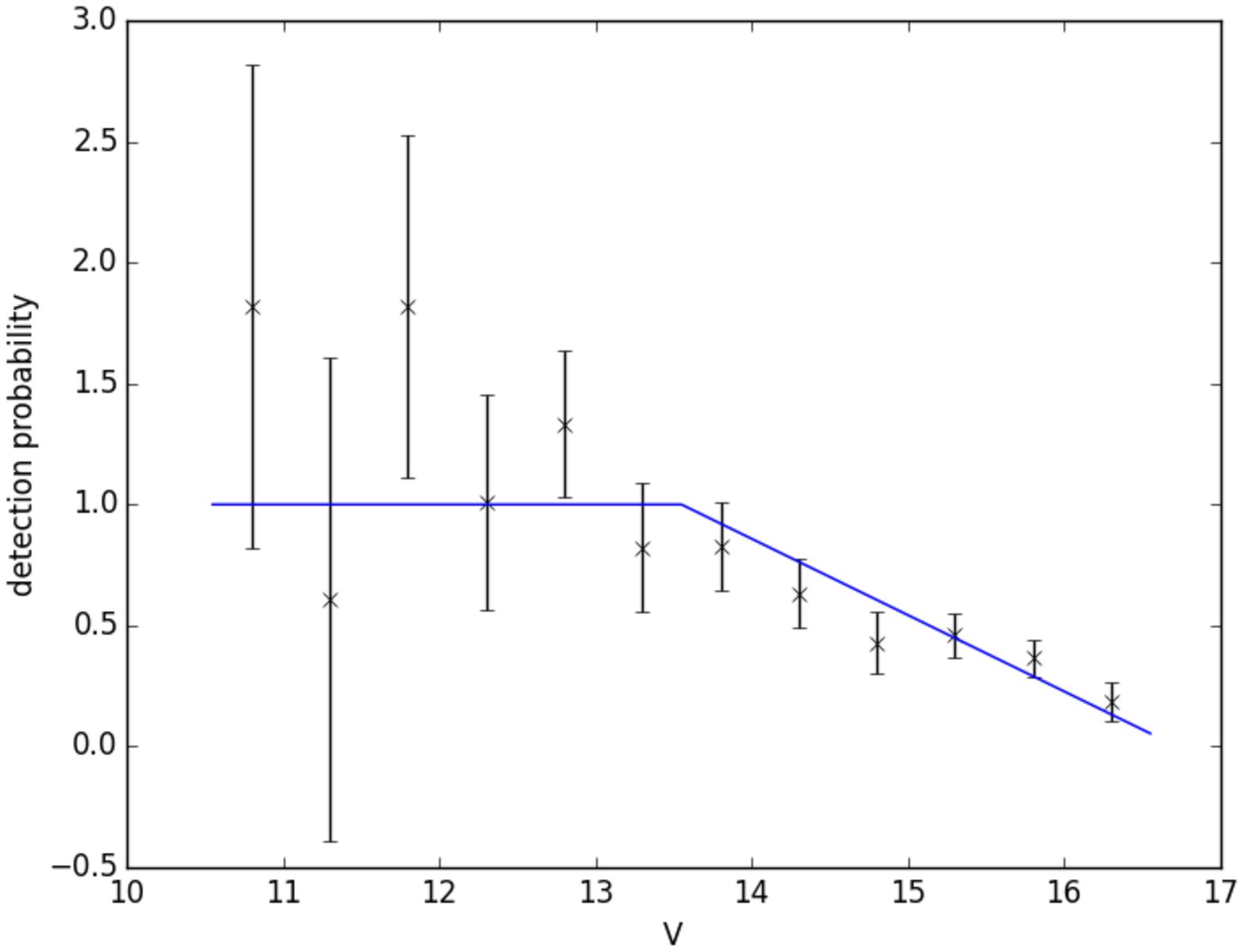}
\caption{Relative probability of a white dwarf being included in SPY as a function of brightness. Error bars are derived from the Poisson distribution. \label{f:detect}}
\end{figure}

Twenty of the 39 binaries found in SPY are double-lined, an
astonishingly large fraction. As remarked above, double-lined systems
are potentially up to $0.75$ magnitudes brighter than single white dwarfs of the same mass, boosting the chance that they are included in a magnitude-limited
sample. Since the SPY sample was compiled from a number of catalogues and
surveys with different faint limits, the survey does not have a well defined
cut-off magnitude, however we can estimate the significance of the effect from the systems in hand.  Figure~\ref{f:detect} shows the probability of a white dwarf to be 
included in SPY. It was derived by comparing the brightness distribution of observed white dwarfs with a population produced using the model described in \citet{2009JPhCS.172a2004N}. 
Numbers of observed white dwarfs found in each bin are compared to the model predictions. Only
relative probabilities are important. The probability is modeled with a simple
linear function and assumed to be one for white dwarfs brighter than 13.55. The fitted
probability $p$ is
\begin{equation}
p = \left\{ \begin{array}{ccl}
1.00           && V\le 13.55\\
5.27-0.3154V   && 13.55\le V \le 16.55\\
0.00           && V>16.55\\
\end{array} \right.
\label{e:prob}
\end{equation}
Fundamental parameters were determined by fitting the spectra with model
spectra representing the two components (see appendix~\ref{s:fundamental} for
details). We then calculated the contribution of the secondary star to the
combined brightness and from that the brightness of the primary star alone. The
probabilities of the system and the primary star alone were calculated using
Eq.~(\ref{e:prob}) and used to estimate the probability of the primary alone
making it into the SPY sample. Example, if the probabilities are 60\% and
30\%, respectively, we count the system as 0.5. It turned out that 12.5 of the
20 double-lined binaries (63\%) would have appeared in the survey. The total number of detected binaries would have dropped from 39 to 31.5 (81\%). 

\subsection{Implications for SN\,Ia progenitor scenarios.}
Since the initiation of the SPY project, ideas about possible progenitors of
SNe Ia have evolved considerably. In particular, the extensive work in
following up SNe Ia has revealed a richer diversity than was previously
imagined \citep{2017hsn..book..317T}. The most profound change perhaps has been a move away from the
Chandrasekhar mass model, as there now seems to be good evidence for
sub-Chandrasekhar ejecta masses in observed SNe\,Ia \citep[e.g., ][]{2014MNRAS.440.1498S}.

We note that among the objects already discovered by SPY there are two systems -- WD2020$-$425 (M=(0.81+0.54)\,\msun), P=0.3 day) and HE2209$-$1444 (M=(0.72+0.43)\,\msun), P=0.28 day), which may well
qualify as candidates for sub-Chandrasekhar SN\,Ias progenitors. Moreover, the SPY project is not completed,
as yet. Out of 39 DDs identified in the survey, only 19 have measured $M_2$ or lower
limits thereof from radial velocity curves. There are 10 SB2 DD still needing follow-up observations by UVES
(see Tab.~\ref{t:RVDD}). The most interesting SB2 system to follow up is
%are WD0128-387, MCT0136$-$2010, and 
HE0324$-$1942, because its primary mass ($\approx$0.8 M$_\odot$) is as high as that of WD2020$-$425.   

The expectation of finding a super-Chandrasekhar pre-SN\,Ia white dwarf pair in the sample of about 1000
objects brighter than $V\approx$\magpt {16}{5}
was based on a count of the number of model systems satisfying the criteria
 $M_{tot}>M_{Ch}$ and $t_{merger}<t_{Hubble}$, while any selection effects were ignored. 
However, a numerical experiment
modelling observations of a similar sample of white dwarfs by an instrument like VLT/UVES in which the effects of observational
selection (orbital inclination, observational strategy, observing conditions, etc.)
were taken into account \citep{2019MNRAS.482.3656R}, has shown that finding such a pair of white dwarfs
may require a much larger sample of objects -- they estimate a factor of ten.
Given that the sample size of 643 presented in the paper does
not match even the original aim of around 1000 targets (owing in the main to
subdwarfs \citep{2005A+A...430..223L,2007A+A...462..269S} which are hard to distinguish from white dwarfs from colours alone,
and not to mind the $\sim 20$\% detection efficiency loss), the SPY sample may
well not deliver a super-Chandrasekhar, short period DD, but with the switch
of emphasis towards sub-Chandrasekhar mergers, it nonetheless retains its
significance, and is the only large survey sensitive to masses and periods typical of the bulk of the DD population.

The discovery of the DDs presented here is only a first step;
the next is the determination of their orbital periods, which has been
accomplished already in some cases, and should be a priority in future work on
this sample. Once this is accomplished, we should be in a position to be able
to compare the properties of DDs in samples generated through binary
population synthesis to an observed sample of relatively weak mass bias, and
hence be able to judge on the SNe Ia rate that the Galactic population of DDs can
plausibly sustain.

\section{Summary}

In this paper we have presented the methods and results of radial velocity
measurements of 643 DA white dwarfs observed with UVES and the VLT as part of
the SPY survey. We find 39 double degenerates in this sample, including 20
double-lined systems which show spectral components from both white
dwarfs. One of the double-lined systems (WD\,2020$-$425) is a candidate SN\,Ia
progenitor with a total mass close to the Chandrasekhar limit, but the uncertainties of the mass for this binary are too large to be certain of this
at present. The absolute radial velocities we presented will allow
for future searches for long-period binarity, while the sample of DDs is
well-suited for tests of binary population models.

The overall DD fraction in the SPY sample is 6\%. We compared our results to the results of \citet{2017MNRAS.467.1414M}, who studied a subsample of the SPY data set and derived a DD fraction of 10\%$\pm$2\% (1$\sigma$, random) +2\% (systematic). We showed that their detection criterion for radial velocity variability is overly optimistic. Applying our detection criterion, we showed that 17 of their candidate DDs are unlikely to be RV variable, reducing their SPY subsample to 25 DDs. Accordingly, the DD fraction in the \citet{2017MNRAS.467.1414M} subsample is in fact 6\%, consistent as it should be with our result for the full SPY sample.

Our detected DD fraction among C/O white dwarfs is $3.9\pm0.8$\%, 
below the 10\% fraction previously found, perhaps reflecting the prominence of low mass He white dwarfs in literature studies. The detected DD fraction amongst He white dwarfs in SPY
is much higher at $36.4\pm7.3$\%, although it seems unlikely that all He white dwarfs are binary. We estimated that the number of binaries needs to be
corrected by a factor of 0.8 to account for the increased probability of double-lined systems to be included in SPY, but there is a similar factor in reverse because our detection efficiency is $80$ to $90$\% for typical DD
periods, and a full comparison with theoretical models needs to take multiple selection effects into account, despite their being weaker in SPY than in any comparable survey. We provide the data needed to accomplish such a task.

\begin{acknowledgements}
  We express our gratitude to the ESO staff, for providing invaluable help and
  conducting the service observations and pipeline reductions, which have made
  this work possible.  TRM and SC were supported under grants from the UK’s
  Science and Technology Facilities Council (STFC), ST/L00073 and ST/P000495/1.
  LRY is partially supported by Presidium of the Russian Academy of Sciences program P-28 and RFBR grant 19-02-00790.  Work on the SPY
  project in Bamberg is supported by the DFG (grants Na365/2-1, Na365/2-2,
  HE1356/40-3 and HE1356/40-4). We thank Markus Dimpel for obtaining the Gaia IDs.
\end{acknowledgements}

\bibliographystyle{aa}
%\bibliography{spy_18,add_spy_lry}
\bibliography{spy_lit}

\begin{appendix}

\section{A test of the possible impact of pressure broadening on Balmer lines}
\label{s:pressure}

The densities in atmospheres of white dwarfs are sufficiently high that
pressure shifts of line profiles are a concern. If present, these effects
mimic an RV shift. Laboratory experiments by \citet{1971ApJ...166L..59W} and
\citet{1972PhRvA...6.1132W} measured substantial pressure redshifts for
\hbeta, H\ensuremath{\gamma}, H\ensuremath{\delta} and to a lesser
extent for \halpha. \citet{1976ApJ...209..205S} included the shifts determined
in these experiments and solved the radiation transfer for white dwarf
atmospheres. The resulting line profiles were analysed following the procedure
applied to observed spectra at that time to determine the line centre. This
investigation found that effects for \halpha and \hbeta were below the
1\,\kms\ level. A later analysis by \citet{1987ApJ...313..750G} revisited the
\citet{1972PhRvA...6.1132W} data and carried out a more detailed fit of the
measurements with three parameters (plus a correction term). Stellar line
profiles were computed using model atmospheres and analysed in a fashion
similar to the previous investigation by \citet{1976ApJ...209..205S}. The new
study by \citet{1987ApJ...313..750G} confirmed small pressure shifts for
\halpha exceeding 1--2\,\kms\ only for cool, high gravity ($\log g = 9$) white
dwarfs. However, for \hbeta the authors find large shifts often in the range
5--10\,\kms.

A recent study by \citet{2015ApJ...808..131H} calculated theoretical line
profiles using a so-called modified full computer simulation method. This
method treats shifts and asymmetries as corrections depending on the local
plasma conditions to a simpler first order description of line profiles.The
authors compare the new calculations with laboratory data and find
improvements to previous attempts. Atmospheric line profiles of \halpha and
\hbeta were calculated using the new theoretical calculations. The result is
even smaller pressure shifts for \halpha than reported in
\citet{1987ApJ...313..750G}. The effect does not exceed 1\,\kms\ even when the
measurement is carried out 25\,\AA\ from the line centre. Contrary to that the
new study finds an effect for \hbeta about twice as large as previously,
exceeding 10\,\kms\ at a distance of 15\,\AA. The authors point out that the
impact on RV measurements should be very small, if the resolved NLTE core of
the lines is measured in high resolution spectra. The Doppler core is formed
in low density regions of the atmosphere where the Stark effect is of no
importance compared to Doppler broadening.

We carried out an empirical test of the possible impact of \hbeta pressure
shifts on our RV measurements. We selected a subset of white dwarfs from Table~\ref{t:RV} with clearly visible NLTE core and good spectra, yielding
an RV accuracy of 2\,\kms or better. We excluded targets with a gravity below
7.8. Lower gravity white dwarfs have lower density atmospheres, which would result in
smaller effects. We selected the first 100 objects fulfilling these criteria
from our catalogue with a total of 201 spectra between them
(high S/N~100; H100).

\begin{table}
\caption{Systematic shifts measured for the H100 sample\label{t:shifts}}
\begin{tabular}{lccr}
measurement&interval&remark&RV shift\\
&\AA&&\kms\\ \hline
RV(\hbeta)--RV(\halpha)  &$\pm 15$\,\AA &\rule{0mm}{4.0mm}   &$-0.8\pm 0.4$\\
RV(\hbeta)--RV(\halpha)  &$\pm 30$\,\AA &   &$-2.5\pm 0.4$\\
RV(\hbeta)--RV(\halpha)  &$\pm 30$\,\AA &core smoothed   &$-8.7\pm 0.6$\\
RV(\halpha+\hbeta)--RV$_{\mathrm{best}}$  &$\pm 30$\,\AA &core smoothed   
                      &$-3.7\pm 0.3$\\
\end{tabular}
\tablefoot{The $1\sigma$ uncertainty of the average is given, calculated using
  bootstrapping.}
\end{table}

Our measurements of the H100 sample are summarised in Table~\ref{t:shifts}.
We first determined the offsets between \halpha and \hbeta measured over the
$\pm 15$\,\AA\ interval, i.e.\ in the same fashion as our original
measurements, only that we then included both lines in a simultaneous fit. The
result is a very small systematic difference between both lines of
$-0.8\,\kms$. The offset increased to $-2.5\,\kms$ when we increased the
fitted interval to $\pm 30$\,\AA. It can be expected that the largest signal
is still delivered by the resolved Doppler broadened NLTE core. We thus
convolved the model spectra with a Gaussian of 3\,\AA\ FWHM, which smoothes
the core away and tweaked the fit process to give small weights to the
Doppler core. This procedure now gives the results from measuring the Stark
broadened wings only. The offset between \hbeta and \halpha increases to
$-8.7\,\kms$.

Somewhat surprisingly we measure on average smaller RVs for \hbeta than for
\halpha, i.e.\ blueshifts not redshifts as predicted by
\citet{2015ApJ...808..131H}. However, let us keep in mind that the fit process
includes a normalisation of model and observed spectrum. It is well possible
that this will cancel out, or maybe, even overcompensate line
asymmetries. \citet{2015ApJ...808..131H} carried out a normalisation of their
model spectra, but used the true continuum for this exercise. The
\fitsb\ normalisation process, as applied for our measurements, restricts
itself to the fitted region. An asymmetry in the outer wings of the line, if
present, will thus result in a tilt. 

A detailed investigation is beyond the scope of this article. Our main interest
here is an estimate of possible systematic offsets of our RVs. The small RV
offset between \halpha and \hbeta measurements for our H100 sample done the
standard way over the $\pm 15$\,\AA\ interval gives us confidence that the
systematic RV errors for white dwarfs with well developed NLTE cores is not larger than
1\,\kms. Note also the good agreement between our RVs measured from
Balmer lines and the Ca\,K line (Fig.~\ref{f:CaRV}). 

We used the best guess RVs (RV$_{\mathrm{best}}$) from Table~\ref{t:RV} as
reference point to get an estimate of systematic offsets for RVs measured
using the $\pm 30$\,\AA\ interval. We fitted RVs of our H100 sample over $\pm
30$\,\AA, combining \halpha and \hbeta, the NLTE core smoothed away. The
average difference to RV$_{\mathrm{best}}$ is $-3.7\,\kms$.

We actually do not know RV$_{\mathrm{best}}$ for DA targets with weak NLTE
cores, for which we needed to use a wider fit interval. However, we made an
attempt to measure the differences between \halpha and \hbeta
RVs. Uncertainties are quite substantial due to the weak features of these
spectra. These measurements also produced a few outliers, a tendency which is
much reduced if both Balmer lines are combined. This is the main reason we included
\hbeta for our RV measurements in the first place. To avoid spurious results
we excluded spectra with an uncertainty of the RV difference exceeding
15\,\kms. The average of the remaining 31 measurements has an RV offset of
$-14.6\pm 2.6\,\kms$. This is larger than what we measured for our H100 sample
with smoothed NLTE core. It is plausible that the generally weaker features of
these stars increase the impact of line asymmetries or other
systematics. Assuming that the pattern observed for the H100 sample holds, we
estimate that the systematic error for the simultaneous fit of \halpha and
\hbeta amounts to about $-7\,\kms$.

In summary we estimate that systematic offsets of the DAs with prominent NLTE
cores (the vast majority) should not exceed 1\,\kms, for targets with weak
Doppler cores a systematic offset of about 7\,\kms is possible. These are
small enough that they will have no impact on the {\it relative} RV
measurements at different epochs, but need to be considered for {\it absolute}
RVs.
 
\section{Radial velocity measurements for all DA white dwarfs}
\label{s:RVtable}

Table~\ref{t:RV} lists all RV measurements of the DA white dwarfs in the SPY
sample.

\paragraph{Column 1} gives the designation. "WD" indicates that this source was
selected from the \citet{1999ApJS..121....1M} version of the WD
catalogue. Other designations indicate that other sources were used (see
text). 

\paragraph{Column 2.} Julian date. The dates are for mid-exposure and
corrected to heliocentric values.

\paragraph{Column 3.} Individual heliocentric RV measurements for the
epochs. If photospheric Ca~II lines were present, the Ca\,K lines was
included in the measurement. Observations done under poor conditions (usually
repeated afterwards) are included in this table. Errors are often large due to
poor signal to noise levels, but could carry useful phase information.

\paragraph{Column 4.} Wavelengths range used for the RV measurements, shifted
according to the measured RV. An empty entry indicates the standard
$\pm$15\,\AA\ interval. Larger values were used, when a sharp line core was
missing (see text). Sometimes an asymmetric range was chosen to avoid flaws in
the spectra.
 
\paragraph{Column 5.} Weighted average of the RV measurements. The error is
computed including the estimate for instrumental systematics calculated from
the telluric lines and comparison between Ca and H lines when present.

\paragraph{Column 6.} $\chi^2$ values computed for the null hypothesis that 
RV is constant. Again, our estimates for the systematic errors are included.

\paragraph{Column 7.} The next column gives the probability $p$ that $\chi^2$ 
has a value as high as the measured one or higher for an RV constant star. The
smaller the values the higher the confidence that the star is RV variable.

\paragraph{Column 8.} A binary is assumed for probability $\log p<-4$ and
marked as DD in this column. Double-lined systems not exceeding our RV variation threshold, but identified by visual inspection are marked dd.

\paragraph{Column 9.} Comments and observations. Abbreviations:
\begin{description}
\item[SB2:] Double-lined DD.
\item[G\&L:] Model spectrum modified using one or more Gaussians and
  Lorentzians (see text). 
\item[polyn.:] Model spectrum modified using a low order polynomial.
\item[$\vrot$:] Rotation included in fit. That does not necessary indicate
  that the star is rotating. The aim was to achieve a good match of the
  observed profiles for fitting RVs without considerations of the physical
  soundness. 
\item[+Ca\,K:] Photospheric Ca~II present and included in fit.
\end{description}

\clearpage

\onecolumn
\begin{longtable}{lrr@{$\pm$}lc|r@{$\pm$}lrrc|l}
%\begin{longtable}{lrrr|rr@{$\pm$}lr|l}
\caption{RV measurements for the SPY DA white dwarfs. \label{t:RV}}\\

\hline\hline
\rule{0mm}{2.6ex}
object &\multicolumn{1}{c}{HJD} &\multicolumn{2}{c}{RV} &$\Delta \lambda$ 
&\multicolumn{2}{c}{$\overline{RV}$} &$\chi^2$ &$\log p$&
&comments\\
 & $-2\,400\,000$ &\multicolumn{2}{c}{\kms} &\AA 
&\multicolumn{2}{c}{\kms} & &     &
\\ \hline
\endfirsthead
\caption{RV measurements, continued.}\\
\hline \hline
\rule{0mm}{2.6ex}
object &\multicolumn{1}{c}{HJD} &\multicolumn{2}{c}{RV} &$\Delta \lambda$ 
&\multicolumn{2}{c}{$\overline{RV}$} &$\chi^2$ &$\log p$&
&comments\\
 & $-2\,400\,000$ &\multicolumn{2}{c}{\kms} &\AA 
&\multicolumn{2}{c}{\kms} & &     &
\\ \hline
\endhead
 WD2359$-$434 &51739.7790 &  49.6 &  1.4 &             &  48.7 &  1.1 &   0.40 & $-$0.28 &   &\poly , \ggll , \vrot\\
      &51743.8041 &  47.7 &  1.5 &             &\multicolumn{2}{c}{ } &        &       &   & \\ \hline
 WD2359$-$324 &51740.7447 &  94.7 &  5.6 &             &  81.6 &  2.4 &   4.36 & $-$0.95 &   &\\
      &51742.8050 &  72.6 &  6.8&             &\multicolumn{2}{c}{ } &        &       &   & \\
      &52799.9168 &  79.9 &  2.7 &             &\multicolumn{2}{c}{ } &        &       &   & \\ \hline
 WD0000$-$186 &51803.7127 &  25.3 &  1.7 &             &  24.2 &  1.3 &   0.36 & $-$0.26 &   &\\
      &51804.6533 &  23.3 &  1.7 &             &\multicolumn{2}{c}{ } &        &       &   & \\ \hline
 HS0002+1635 &52610.5530 &  18.2 &  2.1 &             &  16.4 &  1.4 &   0.63 & $-$0.37 &   &\\
\hline

\end{longtable}

\twocolumn

\section{Fundamental parameters}
\label{s:fundamental}

Table~\ref{t:tefflogg} lists the fundamental parameters surface temperature
and gravity adopted for our targets. Most of these values are taken from
\citet{2009A+A...505..441K} who determined temperatures and gravities of sample
stars by fitting the Balmer lines of the SPY spectra. The table lists effective temperatures and gravities, with the 3D atmosphere corrections of
\citet{2013A+A...552A..13T} for DAs with a convective envelope. White dwarf masses were estimated from interpolation in the white dwarf cooling tracks of
\cite{2007MNRAS.382..779P}, \cite{2010ApJ...717..183R} and \cite{2005A+A...441..689A}.
%\cite{2007A+A...465..249A}.

\citet{2009A+A...505..441K} did not derive parameters for some DAs of the
sample for various reasons, including a few cases of objects overlooked. For
some other cases we decided to replace the values as discussed below. Notes
and references have been included in Column~5 of Table~\ref{t:tefflogg} and
explained in Table~\ref{t:TG_ex}, whenever values different from those given
by \citet{2009A+A...505..441K} were included.

\paragraph{Double-lined binaries:} Light of both components is seen in the 
spectra of double-lined binaries. A deconvolution of both components was
attempted using \fitsb and the same DA model grid as used by
\citet{2009A+A...505..441K}. An interative procedure was carried out starting
with a simple by-eye estimate of the RVs in the spectra, followed by a fit of
temperatures and gravities of both components. These were then used for
improved estimates of RVs from the cores of \halpha and \hbeta as described in
\ref{s:fit_process}. The latter two steps were repeated until the iteration
came to a standstill. The relative contributions of both components were
determined using mass-radius relations from the cooling tracks above. The
fundamental parameters listed in Tables~\ref{t:tefflogg} and~\ref{t:RV}
correspond to the primary (i.e.\ the brighter component).

The fundamental parameters listed for the primaries are tentative and should
be replaced by results from full orbital solutions once these are
available. They were derived for the purpose of assessing the status of the
primary as low-mass He core or C/O core white dwarf.

\paragraph{Magnetic white dwarfs:} Magnetic white dwarfs included in this 
study have fields strong enough to induce a detectable Zeeman splitting in the
cores of the Balmer lines, but the lines do not show distortion beyond that. A
detailed analysis of the field structure was beyond the scope of this article
and we adopted a crude approximation to derive fundamental parameters. Model
spectra from standard model atmospheres were co-added after applying
shifts. The $\pi$ component was assumed unshifted and a symmetric shift was
applied to the two $\sigma$ components. Geometric effects were only taken into
account by adjusting the relative strength of the $\pi$ and $\sigma$
components. This approach is oversimplistic and ignores subtle effects of
radiation transfer and field geometry, but our main aim was to get still
useful estimates of fundamental parameters, even if of somewhat limited
accuracy.

Grids were constructed for a range of splits. For a given magnetic white dwarf
best fit values of temperature and gravity were derived for each grid. Values
corresponding to the Zeeman split yielding the smallest value of $\chi^2$ were
adopted. While we deem the accuracy of the fundamental parameters sufficient
to assess the nature (high or low mass) of the white dwarfs, investigators with
a specific interest in these stars are advised to carry out a more
sophisticated analysis.

\paragraph{Cool white dwarfs:} It has been known for some time that the accuracy of surface gravity measurements 
from Balmer lines alone decreases considerably for white dwarfs at
temperatures below $\approx$6500\,K \citep[see, e.g., Fig.~2 of ][]{2012MNRAS.425.1394K}. Higher accuracy can
be achieved, if parallax measurements are available. We replaced spectroscopy-only values by the parallax based photometric values of
\citet{2012ApJS..199...29G}, if available. Spectroscopic-only fundamental
parameters were not included when assessing the number of single and binary
white dwarfs in the mass ranges in Table~\ref{t:mass_groups} and the discussion
(Sect.~\ref{s:discussion}). This is indicated by the lack of a mass entry in
Table~\ref{t:tefflogg}. All white dwarfs are included in the numbers for the
full sample.

\begin{table}
\caption{References and explanations for Table~\ref{t:tefflogg}}
\label{t:TG_ex}
\begin{tabular}{lp{6cm}}
ambiguous &\citet{2009A+A...505..441K} reported a cool and a hot solution. The
solution yielding the smallest $\chi^2$ is included.\\

BL02&\citet{2002ApJ...566.1091B}  \\

GBD12, phot &\citet{2012ApJS..199...29G}, \teff\ and $\log g$ derived from
photometric spectral energy distribution and trigonometric parallax.\\ 

GBR11 &\citet{2011ApJ...743..138G}, fit of Balmer lines\\

KNV05 &\citet{2005A+A...439..317K}, helium-rich DABZ\\

KV12  &\citet{2012MNRAS.425.1394K}, fit of Balmer lines\\

LBSS93&\citet{1993ApJ...418..426L}, binary of magnetic and non-magnetic WD\\

magnetic & magnetic DAs, Zeeman splitting of the line cores. See text for
details on the fit procedure.\\

new &white dwarf not included in the \citet{2009A+A...505..441K}
analysis. Analysed using the same model grid.\\

new/NLTE &hot DAO white dwarfs, analysed using the grid of NLTE model
atmospheres described in \citet{1999A+A...350..101N}.\\

primary & Parameters of the primary from a simultaneous fit of primary and
secondary. See text.\\

WB94 &\citet{1994ApJ...429..369W}, fit of primary and secondary parameters
using lines and spectral energy distribution\\

HNLE97 & \cite{1997A+A...324L..53H}: genuine DAB, helium line profile variations\\

K09 &\citet{2009A+A...505..441K}: visual double \\

FK97 & \citet{1997ApJ...489L..79F}: visual double\\

J98 & \citet{1998A+A...330..277J}: visual double\\

\end{tabular}
\end{table}

\onecolumn
\begin{longtable}{llrrrlccc}
%\begin{longtable}{lrrr|rr@{$\pm$}lr|l}
\caption{Gaia DR2 IDs, effective temperatures, gravities and masses for the SPY DA white dwarfs. 
\label{t:tefflogg}}\\

\hline\hline \rule{0mm}{2.6ex} 
object & Gaia DR2 ID &\multicolumn{1}{c}{\teff} &\multicolumn{1}{c}{$\log g$} 
         &\multicolumn{1}{c}{$M$} &References/remarks\\ 
    &   &\multicolumn{1}{c}{K}  &\multicolumn{1}{c}{cm\,s$^{-1}$}  
       &\multicolumn{1}{c}{\msun} & \\ \hline
\endfirsthead
\caption{Gaia DR2 IDs, effective temperatures, gravities and masses, continued.}\\
\hline \hline
\rule{0mm}{2.6ex}
object & Gaia DR2 ID &\multicolumn{1}{c}{\teff} &\multicolumn{1}{c}{$\log g$} 
       &\multicolumn{1}{c}{$M$} &References/remarks\\ 
     & &\multicolumn{1}{c}{K}  &\multicolumn{1}{c}{cm\,s$^{-1}$}  
      &\multicolumn{1}{c}{\msun} & \\ \hline
\endhead
%\input {AA_Teff_logg.tab}
% WD &Gaia DR 2 ID&Teff&logg&mass&Ref&
WD2359$-$434&  4994877094997259264&8253&8.01&0.604&magnetic&\\
WD2359$-$324&  2313582750735435776&22478&7.74&0.514&&\\
WD0000$-$186&  2414099622710507904&15264&7.85&0.535&&\\
HS0002+1635&  2772855182928578560&25878&7.86&0.568&&\\
WD0005$-$163&  2416481783371550976&11860&7.60&0.435&ambiguous&\\
WD0011+000&  2545505281002947200&9423&7.77&0.492&&\\
WD0013$-$241&  2336189396997071616&18529&7.90&0.568&&\\
WD0016$-$258&  2323704339384503040&10874&7.80&0.507&&\\
\end{longtable}

\end{appendix}

\end{document}